\documentclass[pra,12pt,floatfix]{revtex4}
\usepackage{graphicx}
\usepackage{amsmath}

\begin{document}
\title{On the applicability of jellium model to the
description of alkali clusters}

\author {Anton Matveentsev}
\email[Email address: ]{anton@rpro.ioffe.rssi.ru}
\affiliation{A. F. Ioffe Physical-Technical Institute of
the Russian Academy of Sciences, 
Polytechnicheskaya 26,
St. Petersburg, Russia 194021}

\author {Andrey  Lyalin}
\altaffiliation[Permanent address: ]
{Institute of Physics, St Petersburg State University,
St Petersburg, Petrodvorez, Russia 198504}
\email[Email address: ]{lyalin@th.physik.uni-frankfurt.de}

\author {Ilia A Solov'yov}
\altaffiliation[Permanent address: ]
{A. F. Ioffe Physical-Technical Institute, St. Petersburg, Russia 194021}
\email[Email address: ]{ilia@th.physik.uni-frankfurt.de}

\author {Andrey V Solov'yov}
\altaffiliation[Permanent address: ]
{A. F. Ioffe Physical-Technical Institute, St. Petersburg, Russia 194021}
\email[Email address: ]{solovyov@th.physik.uni-frankfurt.de}

\author {Walter Greiner}
\affiliation{Institut f\"{u}r Theoretische Physik der Universit\"{a}t
Frankfurt am Main, Robert-Mayer 8-10, Frankfurt am Main, Germany 60054}

\begin{abstract}
This work is devoted to the elucidation the applicability
of jellium model to the description of alkali cluster properties on the
basis of comparison the jellium model results with those derived  
from experiment and within {\it ab initio} theoretical framework.  
On the basis of the
Hartree-Fock and  local-density approximation deformed jellium model
we have  calculated  
the binding energies per atom,
ionization potentials, deformation parameters and the optimized values of the
Wigner-Seitz radii for neutral and singly
charged sodium clusters with the number of atoms $N \leq 20$. 
These characteristics are compared with the results
derived from the {\it ab initio} all-electron simulations
of cluster  electronic and ionic structure 
based  on the density functional theory 
as well as on the post Hartree-Fock perturbation theory on many-electron
correlation interaction. The comparison performed demonstrates the great role
of cluster shape deformations in the formation cluster properties and the quite
reasonable level of applicability of the deformed jellium  model.
\end{abstract}



\maketitle

\section{Introduction}
\label{intro}




During the last decade, investigation of the detailed structure and properties 
of small sodium clusters attracted a lot of attention 
(see, e.g., \cite{MetCl99,LesHouches,StructNa} and references therein), 
because namely the sodium clusters were used in such important experimental 
work as the discovery of metal cluster electron shell structure 
\cite{Knight84} and the observation of plasmon resonances
\cite{Br89,Selby89,Selby91}. These experiments were definitely among 
those, which clearly demonstrated that
atomic clusters and small nanoparticles are in fact new
physical objects possessing their own properties.
The novelty of cluster physics is also greatly connected with
the fact that cluster properties explain
the transition from
single atoms or molecules to solid state.
Comprehensive survey of the field can be found 
in review papers and books, see, e.g.,
\cite{deHeer93,Brack93,BrCo94,Haberland94,Guet97,MetCl99,LesHouches}.

With the discovery of electronic shell structure in free
alkali clusters \cite{Knight84,Ekky84} the essential role of
the quantized motion of delocalized valence electrons 
in the mean field created by ions in a cluster has been understood.
Under different experimental conditions, the detailed ionic structure
has been found not to affect the properties of alkali and
other simple metal clusters very much (see, e.g.,
\cite{deHeer93,Brack93,BrCo94} for review).
This behavior suggests the validity of a jellium
model, defined by a Hamiltonian which treats the electrons in the usual 
quantum mechanical way,
but approximates the field of the ionic cores by treating them as a uniform
positively charged background. This naturally leads to a description of the
electron density in terms of single particle wave functions that extend over
the entire cluster.

Initially, jellium calculations for metal clusters
were based on the density functional formalism with the use of pseudopotentials
for the description of electron relaxation effects and lattice structure
\cite{Martins81}. Fully self-consistent calculations for spherical
jellium metal clusters
have been performed within the framework of the spin-density-functional method
\cite{Hintermann83} and the Kohn-Sham formalism
for the self-consistent determination of electron wave functions
\cite{Ekky84,Ekky85}. The Hartree-Fock (HF) scheme for the self-consistent
determination of the electron wave functions of spherical jellium
metal clusters was also introduced later in
\cite{GJ92,IIKZ93}. 

Shortly after the discovery of electronic shell structure in free
alkali clusters \cite{Knight84,Ekky84} it was realized that the detailed
 size dependence of ionization potentials and other characteristics
of small metallic clusters can be understood as a consequence of non-spherical
cluster shapes \cite{Clemenger85} by analogy with the nuclear shell
model. Direct evidence for cluster deformation was achieved in
experiments on photoabsorption, where splitting of plasmon resonances
caused by cluster deformation was observed (see
\cite{deHeer93,Brack93,BrCo94} and references therein).

Kohn-Sham calculations for spherical \cite{Ekky84,Ekky85}
metal clusters have been generalized for spheroidal \cite{Ekky88,Ekky91} and
more general axial shapes \cite{Montag95} of light clusters.
Light clusters of arbitrary shapes have been studied by means of
the "ultimate" jellium model \cite{Koskinen95,Koskinen95a}.
The Hartree-Fock approach was generalized for axially
deformed cluster systems in \cite{LSSCG00,Lyalin01a}.
The existence of the different shape isomers were discussed in
\cite{Montag95,Koskinen95a,Hirschmann93}.
It was shown that the shape of the magic clusters can deviate from the sphere
if higher multipole deformations are taken into account. For example,
the magic cluster $Na_{40}$ is not spherical if octupole deformations
are allowed. This possibility have been pointed out in \cite{Hamamoto91}
and confirmed by the `ultimate' jellium \cite{Reimann97} and the Born-Oppenheimer
local-spin density molecular dynamics method \cite{Rytkonen98}.
It has been shown that alkali-metal clusters have similar shapes
with small atomic nuclei \cite{Hakkinen97}.
This similarity is a universal result of the density-functional theory.


The shell-correction method known from nuclear physics
\cite{Strutinsky67} has been used in a number
of papers in studying spheroidal ground state deformations, 
energetics, stability towards various
fragmentation channels of metal clusters \cite{Yannouleas93a,Yannouleas93b,Reimann93,Bulgac93}. 
The role of higher multipole and tri-axial deformations
has been elucidated in 
 \cite{Lauritsch91,Frauendorf93,Yannouleas95,Frauendorf95,Frauendorf96}.
Using the shell-correction method, the investigation of metal cluster electronic properties
such as binding energies, ionization potentials, electron affinities, energetics of 
fission channels and systematic comparison of the theoretical results with 
the available experimental data have been done in \cite{Yannouleas95} 
(see also \cite{Yannouleas_in_MetCl} and references therein for a review).


Dynamical jellium model for
metal clusters, which treats simultaneously collective vibrational modes
(volume vibrations, i.e. breathing, and shape vibrations) of the ionic jellium
background in a cluster, quantized electron motion and
interaction between the electronic and ionic subsystems
was developed in \cite{GSG99,GISG00}. This model allowed the widths of 
electron excitations in metal clusters beyond the adiabatic 
approximation to be described.
                   
The jellium model provides a very useful basis for studying
various collision processes, such
as photabsorption \cite{Ivanov96}, photoionization
\cite{Bertsch91,C20-C60},
elastic \cite{GCSG97,GEMS98}
and inelastic scattering \cite{GEMS98,GIS97,GISG98,GIPS00},
electron attachment \cite{CGIS98,CGIS99},
photon emission \cite{GS97,GIS98}, 
atomic cluster fission process \cite{LSGS02,LSG02} and others,
involving metal clusters.  On the basis of the
jellium model
one can develop {\it ab initio} many-body theories, such as the
random phase approximation with exchange  or the
Dyson equation method 
and effectively solve many-electron correlation problem
even for relatively large cluster systems containing
up to 100 atoms or even more. Review of these methods in
their application to the electron scattering of metal clusters
one can find in \cite{AVSol}.
As elucidated in the papers cited above,
many-electron correlations are quite
essential for the correct description of various
characteristics of the cluster systems.

Structural properties of small metal clusters have
been widely investigated using quantum chemistry {\it ab initio} methods.
Here we refer to the papers 
\cite{Boustani87,Bonacic88,Boustani88,Guest88,Martins85,Spieg98,
Nagueira99,Gutierrez01,StructNa},
in which optimized geometries, binding energies, ionization potentials, 
electron structure and electron transport properties
of small lithium and sodium clusters have been calculated.

In spite of the fact that both jellium model results and  results
of {\it ab initio} frameworks do exist in literature there have been performed
no systematic comparison of the results of the two different theoretical
schemes so far. We fill this gap in our present paper and demonstrate that
such a comparison is rather illustrative and explains essential physical
aspects of the formation of various cluster characteristics and properties. 
Also, we compare the results of the Hartree-Fock and 
local-density approximation (LDA) deformed jellium 
models and on this basis elucidate the role of many-electron correlation 
effects in the formation of cluster deformations.

On the basis of comparison of the jellium model results with those derived
within the {\it ab initio} theoretical framework and experiment  we  elucidate
the level of applicability of the jellium model to the description of alkali
cluster properties. For neutral and singly charged sodium clusters with
$N \leq 20$, we have  calculated  on the basis of Hartree-Fock
and LDA deformed jellium model the binding energies per atom, 
ionization potentials,
parameters of deformation and optimized values of the Wigner-Seitz radii.
These characteristics are compared with the results derived from 
the {\it ab
initio} all-electron  theoretical framework for the calculation of 
the cluster ionic and electronic
structure based  on the density functional theory as well as on 
the post
Hartree-Fock perturbation theory on many-electron correlation interaction.
Comparison performed in our work demonstrates the great role of cluster
deformations in the formation cluster properties and the quite reasonable level
of applicability of the deformed jellium  model. 

Our paper is organized as follows.  In section \ref{theory}, we provide a
brief review of theoretical approaches and methods used in the calculation.
In section \ref{results}, we present and discuss jellium and 
{\it ab initio} results
and make their comparison. In section \ref{conclusion}, we draw a conclusion
to this paper.

We use the atomic system of units $\hbar=|e|=m_e=1$ in this paper.

\section{Theoretical methods}
\label{theory}

In this work we calculate  
the binding energies per atom,
ionization potentials, deformation parameters and the optimized values of the
Wigner-Seitz radii for neutral and singly
charged sodium clusters with $N \leq 20$
using the Hartree-Fock and LDA deformed jellium model. 
The jellium model results are compared with those derived in \cite{StructNa} 
on the basis of
{\it ab initio} all-electron simulations
of the cluster  electronic and ionic structure.
Below we present a brief review of the theoretical methods used in our
work.
Since the main part of calculations have been done within the 
framework of the deformed jellium model, we focus in our brief review
on the jellium model approach rather than on the details of the
more sophisticated {\it ab initio} methods for which we refer to 
\cite{StructNa}.

\subsection{Two-centered jellium model and cluster shape parameterization}

According to the main postulate of the jellium model, the electron motion
in a metallic cluster takes place in the field of the uniform positive charge
distribution of the ionic background. Originally, the Hartree-Fock model for
metal-cluster electron structure has been worked out in the framework of 
spherically symmetric jellium approximation in \cite{GJ92,IIKZ93}. It is
valid for clusters with closed electronic shells that correspond to magic
numbers (8, 20, 34, 40, 58,...). For metal clusters with arbitrary number of
valence electrons an open-shell two-center jellium Hartree-Fock approximation
has been developed (see \cite{LSSCG00,Lyalin01a,LSGS02}). The two-centered
jellium method treats the quantized electron motion in the field of the
spheroidal ionic jellium background, whose
principle diameters $a$ and $b$ can be expressed as follows:
\begin{equation}
a = \left( \frac{2+\delta}{2-\delta}\right)^{2/3}R, \ \ \ \
b = \left( \frac{2-\delta}{2+\delta}\right)^{1/3}R.
\label{diameters}
\end{equation}

\noindent
$R=r_s N^{1/3}$ is the radius of an undeformed spherical
cluster with $N$ atoms, $r_s$ is the Wigner-Seitz radius,
which for the bulk sodium is equal to 4.0.
The deformation parameter $\delta$ characterizes 
the families of the prolate ($\delta > 0$), 
and the oblate ($\delta < 0$)
spheroids of equal volume $V_c=4\pi a b^{2}/3=4\pi R^3/3$.

The electrostatic potential $U$ of the ionic background can be
determined from the solution of the corresponding Poisson's equation:
\begin{equation}
\Delta U ({\bf r})= - 4\pi \rho({\bf r}),
\label{peq}
\end{equation}

\noindent
where
\begin{equation}
\rho=\begin{cases}
\rho_c, & {(x^2 + y^2)}/b^2 + z^2/a^2 \leq 1 \\
0,      & {(x^2 + y^2)}/b^2 + z^2/a^2 > 1
\end{cases}
\label{rho}
\end{equation}

\noindent
describes a uniform distribution of the ions charge density
in the volume of the cluster. Here $\rho_c= Z_c/V_c$ is the 
ionic charge density inside the cluster, and $Z_c$ is the total charge
of the ionic core.

\subsection{Hartree-Fock and LDA formalism}

The Hartree-Fock equations can be written out explicitly in the form
(see, e.g., \cite{Lindgren})
\begin{equation}
\left( - \Delta/2 + U + U_{HF}\right) \mid a > \, = \,
\varepsilon_a \mid a >.
\label{HF}
\end{equation}

\noindent
The first term here represents the kinetic energy of electron $a$,
and $U$ its attraction to the cluster core. The Hartree-Fock
potential $U_{HF}$ represents the average Coulomb interaction
of electron $a$ with the other electrons in the cluster, including
the non-local exchange interaction,
and $\varepsilon_a$ describes the single electron energy.

According to the density-functional theory,
the ground state energy 
reaches its minimum as a function of the density of the system at
the exact density \cite{Hohenberg-Kohn}.
A self-consistent method for calculation
of the electronic states of many-electron systems was proposed by
Kohn and Sham \cite{Kohn-Sham}. This method leads to the
Kohn-Sham LDA self-consistent equations:
\begin{equation}
\left( - \Delta/2 + U +  U_{H} + V_{xc}\right) \mid a > \, = \,
\varepsilon_a \mid a >.
\label{LDA}
\end{equation}
\noindent

Here $U_{H}$ is the Hartree potential, which represents the direct
Coulomb interaction of electron $a$ with other electrons in the cluster,
but does not take into account the non-local exchange effects, while
$V_{xc}$ is the phenomenological density dependent local
exchange-correlation potential. 
The important feature of the LDA method consist in the fact that it 
takes into account many-electron correlations 
(see, e.g., \cite{Parr-book,Fulde-book} for review).
In the present work we use the Gunnarsson and Lundqvist
model \cite{Gunnarsson}
for the LDA electron exchange-correlation energy density $\epsilon_{xc}$,
which reads as
\begin{equation}
\epsilon_{xc}(\rho_{el}({\bf r})) =
-\frac{3}{4}\left( \frac{9}{4 \pi^2} \right)^{1/3}
\frac{1}{r_s({\bf r})} -
0.0333\ G \left( r_s({\bf r})/11.4 \right).
\label{GunLun}
\end{equation}
\noindent

Here
$r_s({\bf r})=(3/4\pi\rho_{el}({\bf r}))^{1/3}$ is a
{\it local} Wigner-Seitz radius, while $\rho_{el}({\bf r})$
is the electron density in the cluster, and the function $G(x)$
is defined by following relation:
\begin{equation}
G(x)= (1+x^3)\ln \left(1+\frac{1}{x} \right) -
x^2 +  \frac{x}{2} - \frac{1}{3}.
\label{Gx}
\end{equation}

The exchange-correlation energy density $\epsilon_{xc}(\rho_{el}({\bf r}))$,
defines the LDA exchange-correlation potential $V_{xc}(\rho_{el}({\bf r}))$ as 
\begin{eqnarray}
V_{xc}(\rho_{el}({\bf r})) &=& \frac{\delta\left[
\rho_{el}({\bf r})\epsilon_{xc}(\rho_{el}({\bf r}))\right]}
{\delta\rho_{el}({\bf r})} = \\ \nonumber
&& -\left( \frac{9}{4 \pi^2} \right)^{1/3} \frac{1}{r_s({\bf r})} -
0.0333\ \ln\left(1+ \frac{11.4}{r_s({\bf r})} \right).
\label{LDA-pot}
\end{eqnarray}

The Hartree-Fock (\ref{HF}) and LDA (\ref{LDA}) equations
have been solved in the system of the prolate spheroidal coordinates
\cite{KPS} as a system of coupled two-dimensional second order partial differential 
equations. The partial differential equations have been discretized on a
two-dimensional grid and the resulting system of linear equations has been
solved numerically by the successive overrelaxation method \cite{SOR}.
The third dimension, the azimuthal angle has been treated analytically.

An important characteristic of the cluster,
which defines its stability is the total energy $E_{tot}(N,\delta)$.
The total energy depends on the size $N$ of the cluster 
and its core deformation $\delta$.
The total energy $E_{tot}(N,\delta)$ 
is equal to the sum of the electrostatic energy of 
the ionic core
$E_{core}(N,\delta)$  and the energy of the valence electrons
$E_{el}(N,\delta)$:
\begin{equation}
    E_{tot}(N,\delta)=E_{core}(N,\delta) + E_{el}(N,\delta).
\label{eq:E-tot-sum}
\end{equation}

The electrostatic energy of the cluster ionic core is equal to
\begin{equation}
E_{core}(N,\delta) = \frac{1}{2} 
\int_V \rho({\bf r}) U({\bf r}) {\rm d}{\bf r}.
\label{E-core}
\end{equation}

In the HF approximation, the
electronic energy $E_{el}(N,\delta)$ is given by
the general expression \cite{Lindgren}:
\begin{eqnarray}
E^{HF}_{el}(N,\delta) & = &
\sum_{a} < a \mid - \Delta/2 + U \mid a> + \nonumber \\
&& \frac{1}{2} \sum_{abk} q_a q_b
\left[ c(abk) F^{k}(a,b) + d(abk) G^{k}(a,b) \right],
\label{E-el}
\end{eqnarray}

\noindent
where $a$ and $b$ run over all shells. The values
$F^{k}(a,b)$ and $G^{k}(a,b)$ in the Eq. (\ref{E-el})
are the Coulomb and exchange
Slater integrals, $q_a$ and $q_b$ are the occupation numbers
for orbitals $a$ and $b$, respectively. The Hatree-Fock coefficients 
$c(abk)$ and $d(abk)$ for the Coulomb and exchange energy contributions
depend on the occupation numbers (see for details \cite{Lindgren}).

In the framework of LDA the electronic energy of the system 
is given by \cite{Hohenberg-Kohn,Kohn-Sham}: 
\begin{eqnarray}
E^{LDA}_{el}(N,\delta) & = &
\sum_{a} < a \mid - \Delta/2 + U \mid a> + \nonumber \\
&& \frac{1}{2} \int
\frac{\rho_{el}({\bf r}) \rho_{el}({\bf r'})}{{\bf |r-r'|}}
{\rm d}{\bf r} {\rm d}{\bf r'} +
\int \rho_{el}({\bf r}) \epsilon_{xc}(\rho_{el}({\bf r})) {\rm d}{\bf r},
\label{E-el-LDA}
\end{eqnarray}

\noindent where the latter term represents the exchange-correlation energy.

\subsection{All-electron {\it ab initio} Hartree-Fock and LDA calculations}

When performing all-electron {\it ab initio} Hartree-Fock or LDA calculations
for metal clusters,  one has to solve  equations which are symbolically
the same to the Hartree-Fock (\ref{HF}) and the Kohn-Sham (\ref{LDA}) equations 
initially written for the jellium model case.
In {\it ab initio} calculations, instead of using 
the spheroidal jellium parameterization (\ref{rho}) for the distribution of the
ionic charge density and solving the equations only for the valence electrons, 
one has to deal with the exact Coulomb forces of
all the ions
and to solve equations (\ref{HF}) or (\ref{LDA}) assuming that all electrons are
present in the system.
The explicit form of the exchange-correlation density functional
in equations (\ref{LDA}) can be chosen differently. There are many
different functional forms, although there is no unique
one. For example in \cite{StructNa}, the {\it ab initio} results
have been obtained with the use of the
gradient-corrected Becke-type three-parameter exchange functional
\cite{Becke88} paired with the gradient-corrected Lee, Yang and Parr correlation
functional (B3LYP) \cite{LYP,Parr-book}.
The post Hartree-Fock theories accounting
for many-electron correlations, such as for example
the M\o ller-Plesset perturbation theory of the fourth order (MP4) 
\cite{MP} have also been used in \cite{StructNa} for metal 
cluster simulations.

In this paper we do not  present the explicit forms
of these functionals and omit the discussion of the methods used
for the solution of the Hartree-Fock and LDA equations in the {\it ab initio} 
framework. Instead, we refer to the papers, where these functionals
and  methods are presented (see, e.g.,
\cite{StructNa,Becke88,LYP,Gaussian98_man,VWN,PerWan,MP,LL3} 
and references therein).

\section{Numerical results and discussion}
\label{results}
In  this section we  present the results of systematic numerical 
calculations performed for the small sodium clusters in the size 
range $N \leq 20$ on the basis of  deformed Hartree-Fock and  
LDA jellium model.  We determine the binding energies per atom, 
ionization potentials and cluster deformations. These Hartree-Fock and LDA results are 
compared with each other, with the available 
experimental data, and  
with the results derived from 
the {\it ab initio} theoretical framework.  This comparison  elucidates the important role 
of many-electron correlations in metal clusters and  establishes 
the level of applicability of the deformed jellium model.

\subsection{Total cluster energy minimization}

In the Hartree-Fock and LDA jellium models the total cluster 
energy (\ref{E-el}) and (\ref{E-el-LDA})  depends on the
deformation parameter $\delta$ introduced in (\ref{diameters}).

Varying the total cluster energy on $\delta$
one can find its  minimum at certain  $\delta$
for each electron configuration considered. 
This value of $\delta$ corresponds to the stable geometrical
configuration of the cluster.

\begin{figure}
\begin{center}
\includegraphics[scale=1.5]{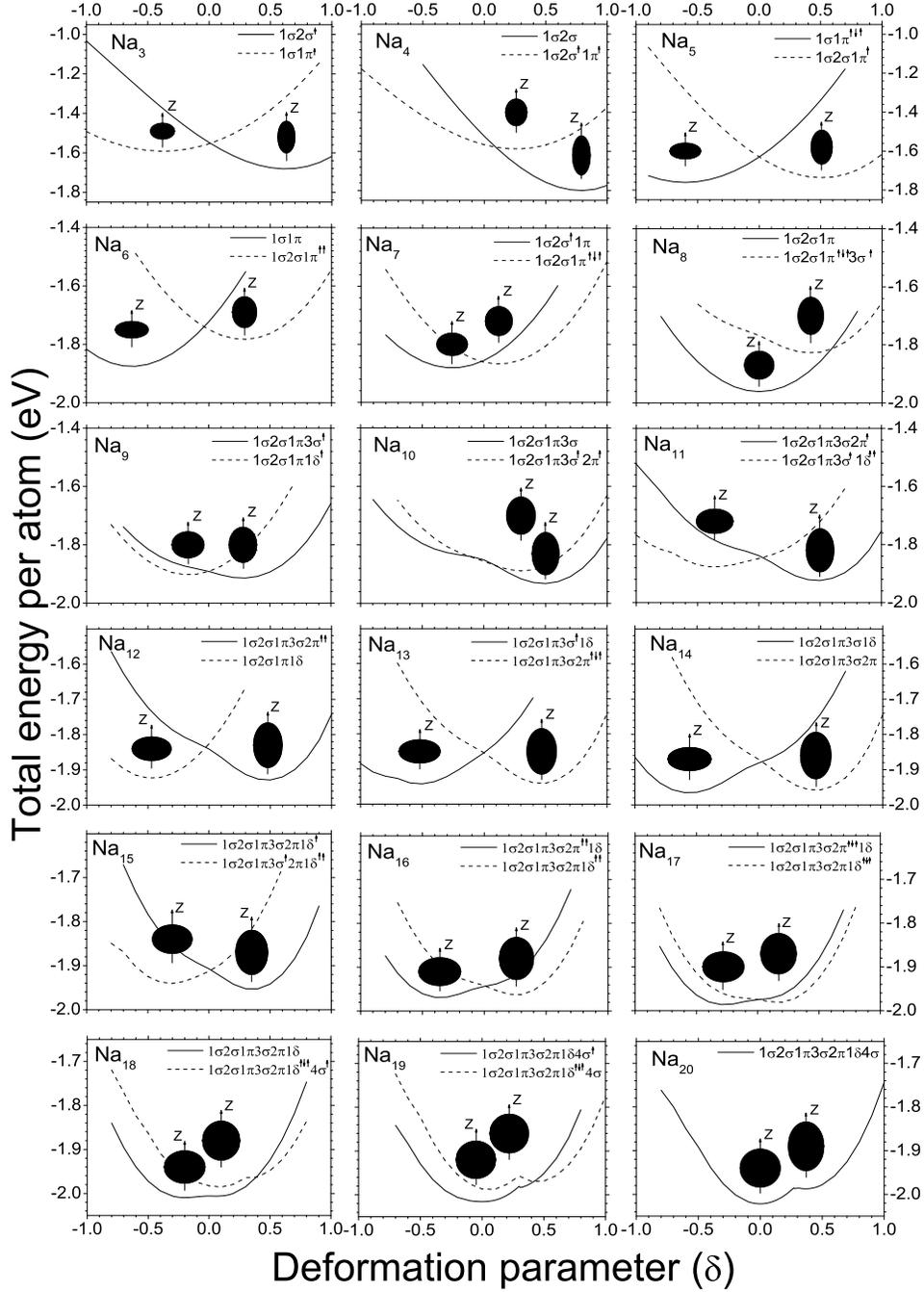}
\end{center}
\caption{Total energies per atom of neutral sodium clusters versus deformation
parameter $\delta$ calculated in the LDA deformed jellium model 
for different electronic configurations.
Solid lines show results for those electronic configurations
of clusters, that result in the absolute minimum of the total cluster energy.
Dashed curves show $E_{tot}(N,\delta)/N$ for the electronic configurations providing
the energy minimum the most closely located to the absolute minimum of the total energy. 
Cluster images represent correctly the relative cluster sizes, as well as type and value 
of cluster deformations.}
\label{tot_en_n}
\end{figure}

\begin{figure}
\begin{center}
\includegraphics[scale=1.5]{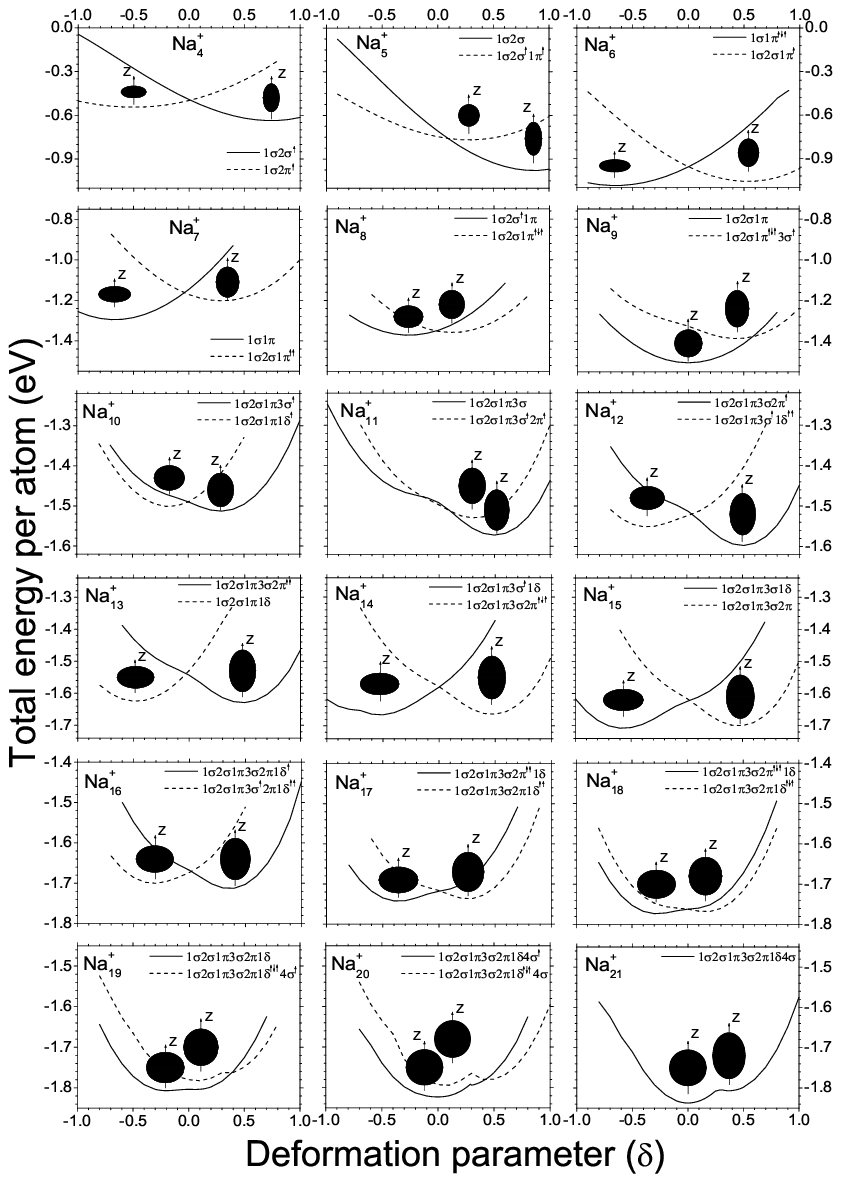}
\end{center}
\caption{The same as Fig. \ref{tot_en_n}, 
but for the singly charged sodium cluster ions.}
\label{tot_en_i}
\end{figure}

In figures \ref{tot_en_n} and \ref{tot_en_i} we present the 
total cluster energy per atom
$E_{tot}(N,\delta)/N$ calculated as a function of deformation parameter $\delta$
for the neutral and singly charged
sodium clusters in the size range $N \leq 20$.

Solid lines in figures \ref{tot_en_n} and \ref{tot_en_i} present 
$E_{tot}(N,\delta)/N$ for those electronic configurations
of clusters, that result in the absolute minimum of the total cluster energy.
The  $\delta$-value corresponding to the minimum of $E_{tot}(N,\delta)/N$ characterizes
the cluster shape at the equilibrium point.
Dashed curves show $E_{tot}(N,\delta)/N$ for the electronic configurations providing
the energy minimum the most closely located to the absolute minimum of the total energy. 
These figures demonstrate that for many clusters with open electron shells
both oblate and prolate isomers are possible and have close energies. The
type of deformation that develops in the cluster is determined
by the type of the corresponding electronic configuration. If the 
electronic orbitals characterizing the chosen electronic configuration 
are alongated with respect to the cluster axis of symmetry then the prolate
deformation of the ionic jellium background is preferable. In the
opposite case, when electronic orbitals lay mostly at the plane 
perpendicular to the cluster axis of symmetry the oblate deformation
of the cluster becomes energetically more favorable. 

It is interesting that the deformed jellium model predicts 
the existence of a non-spherical isomer for the $Na_{20}$ magic cluster. 
So for $Na_{20}$, the second minimum in $E_{tot}(N,\delta)/N$ arises at 
$\delta\approx 0.37$,
while the electronic configuration keeps the same.
In spite of the fact that this second local minimum is energetically 
unfavorable, this prediction is qualitatively correct, because it
corresponds to the result
of {\it ab initio} calculations \cite{StructNa}, which prove
the existence of the two cluster isomers (pyramid and deformed structure
based on the two linked icosahedrons) with the close energies. 
The pyramid cluster isomer possesses
the tetrahedral group of symmetry, which is rather high. Thus, this isomer 
is analogous to the spherical cluster in the jellium picture. The 
deformed icosahedral structure corresponds to the prolate 
jellium cluster configuration. This correspondence  would probably be even 
better if one allows the tri-axial deformations in the jellium approach.

The physical reason for the non-monotonous behaviour of the total energy 
$E_{tot}(N,\delta)/N$ with increasing the deformation parameter  $\delta$
is the strong mixing of the highest occupied $4\sigma$ 
state with the unoccupied  $5\sigma$ state.
These levels exhibit a rather peculiar behaviour with the deformation parameter $\delta$
avoiding each other at $\delta\approx 0.37$. 
Such an avoided crossing-point is linked to the Wigner's no-crossing theorem
\cite{wigner}, which states that two energy levels of the same symmetry
cannot cross. Thus, the strong interaction of the $4\sigma$ state 
with the more prolate type $5\sigma$ state results in the appearance of the 
second minima at $\delta\approx 0.37$ for $Na_{20}$ cluster.

Figure \ref{tot_en_i} demonstrates that similar 
behaviour of the total energy upon deformation parameter observe
for singly
charged cluster ions. This figure shows that 
in the jellium approach the closed shell cluster ions, 
$Na_9^+$ and $Na_{21}^+$, have the spherical shape, $\delta=0$, at the equilibrium
point, similarly to the neutral magic clusters $Na_8$ and $Na_{20}$.

\subsection{Cluster deformations}

In the axially deformed jellium model clusters can either be  spherical or
have a shape of ellipsoid of revolution (spheroid). 
The spheroidal shape can be of the two types prolate or oblate, depending
on the sign of the deformation parameter $\delta$  introduced in 
(\ref{diameters}). In the {\it ab intio} approach, 
the cluster shape is determined by the optimized coordinates of 
all the ions and it can be characterized by the tensor
$R_{ij}$
\begin{equation}
R_{ij}=\sum x_i x_j
\label{tensor_R}
\end{equation}

Here, the summation is performed over all ions in the system. The principle
values of this tensor $R_{xx}$, $R_{yy}$  and $R_{zz}$ define the dimensions
$R_{x}$, $R_{y}$ and $R_{z}$ 
of the ionic charge distribution in the cluster along the
principle axes  $x$, $y$ and $z$.
Note that tensor $R_{ij}$ is
closely connected with the cluster moment of inertia tensor
and the quadrupole moment tensor of the ionic distribution. 

The tensor $R_{ij}$ can also be defined for the jellium model.
In this case, sum in (\ref{tensor_R}) should be replaced
by the integral and the integration to be performed over the
homogeneous spheroidal distribution of the ionic density in the
cluster. Then, the principal values of the tensor $R_{ij}$ can
easily be determined. The result of this calculation reads as

\begin{equation}
R_{xx}= R_{yy}= \frac{b^2}{5}N, \,\,\,\,\,\,\,\,\, 
R_{zz}= \frac{a^2}{5}N. 
\label{tensor_R_j}
\end{equation}

Here, $a$ and $b$ are the principle diameters
of the spheroid defined in (\ref{diameters}).

\begin{figure}
\begin{center}
\includegraphics[bb=15 283 828 851, scale=0.5]{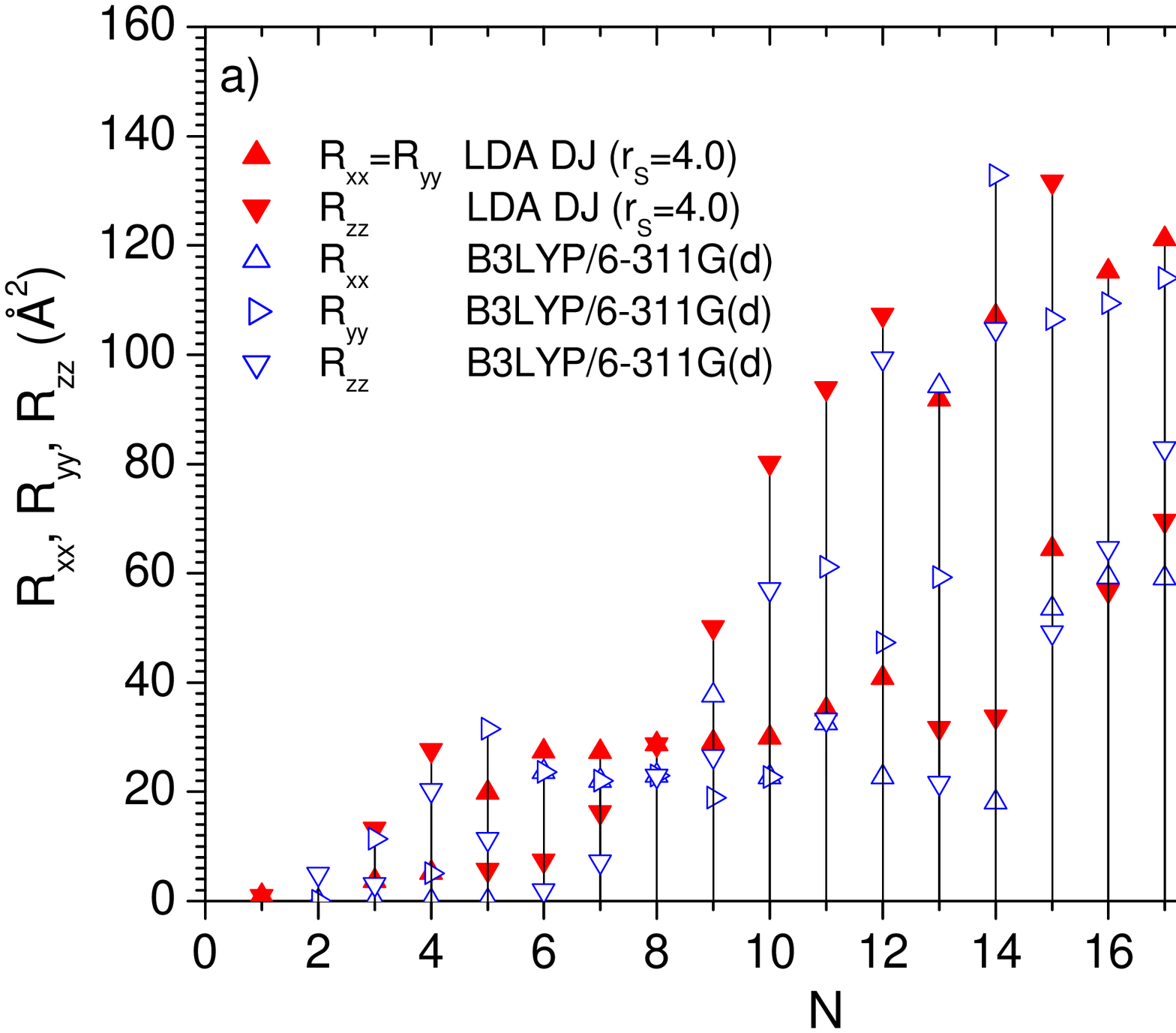}
\includegraphics[bb=15 283 828 851, scale=0.5]{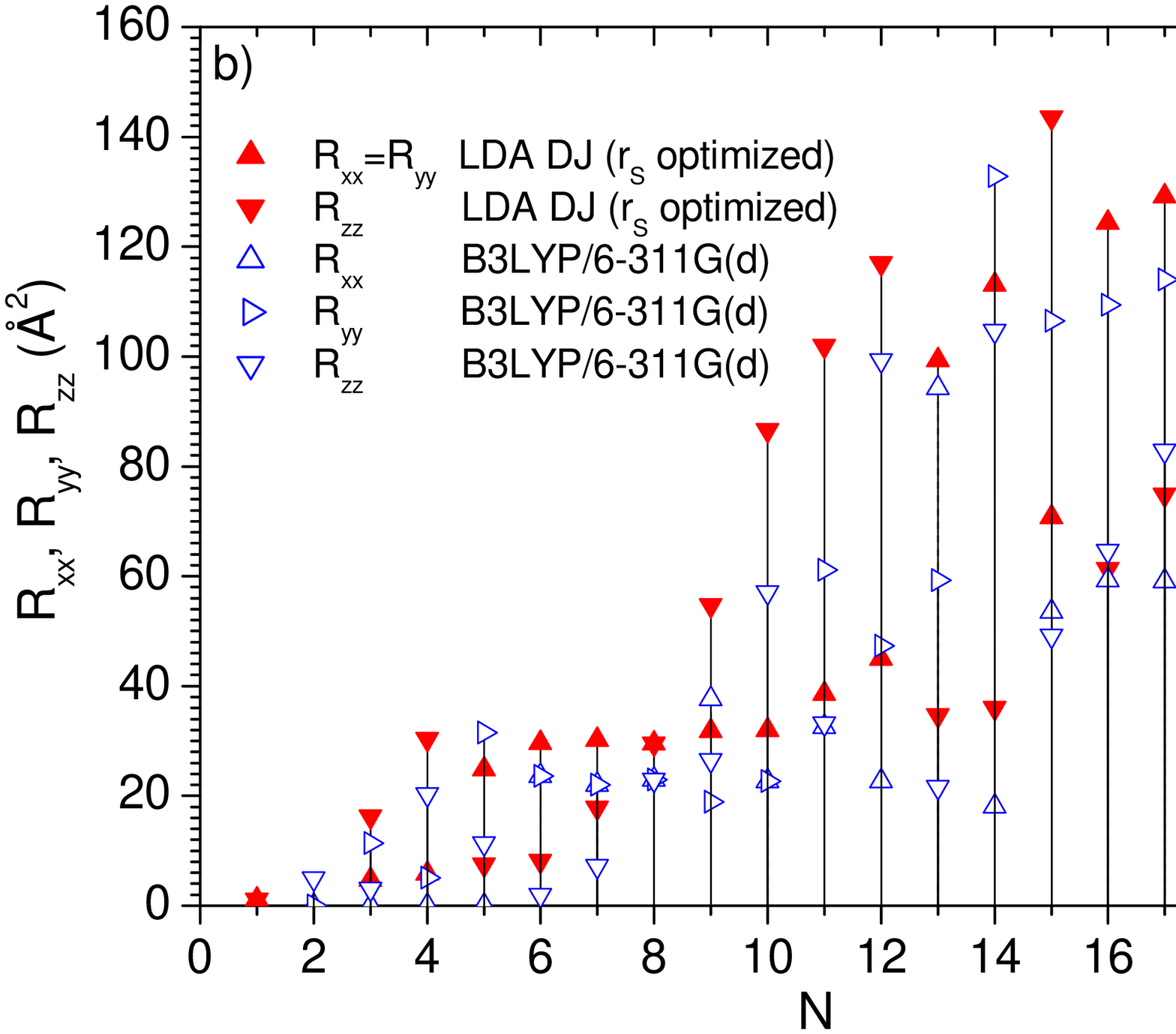}
\end{center}
\caption{The principal values of the tensor $R_{ij}$ for
neutral sodium clusters without Wigner-Seitz radius optimization  
(a) and with optimization on $r_s$ (b) as a function of cluster size
calculated in the LDA deformed jellium (LDA DJ) model (filled triangles) 
and {\it ab initio} B3LYP framework \cite{StructNa} (opened triangles).
}
\label{Tensor_n}
\end{figure}

\begin{figure}
\begin{center}
\includegraphics[bb=15 283 828 851, scale=0.5]{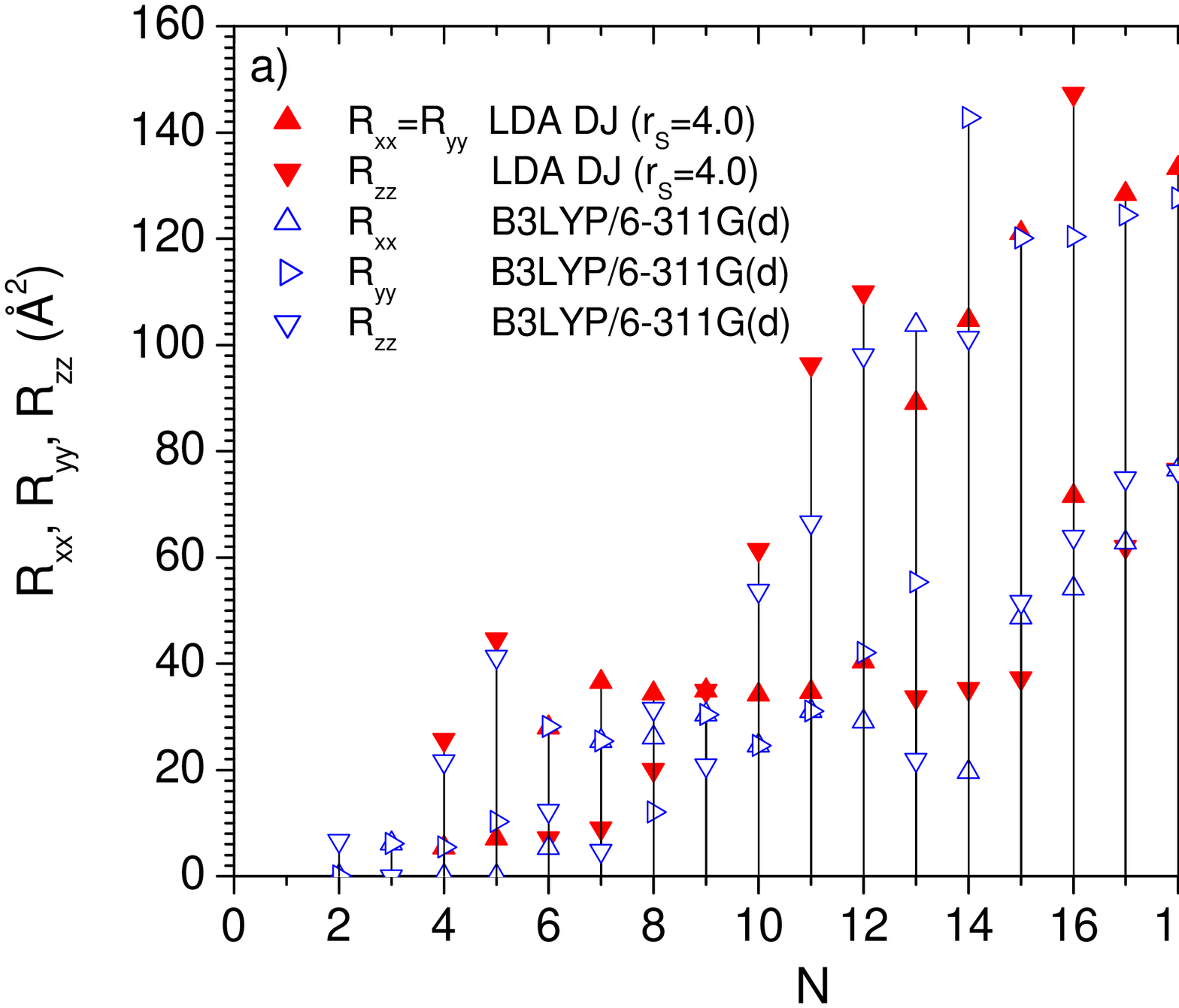}
\includegraphics[bb=15 283 828 851, scale=0.5]{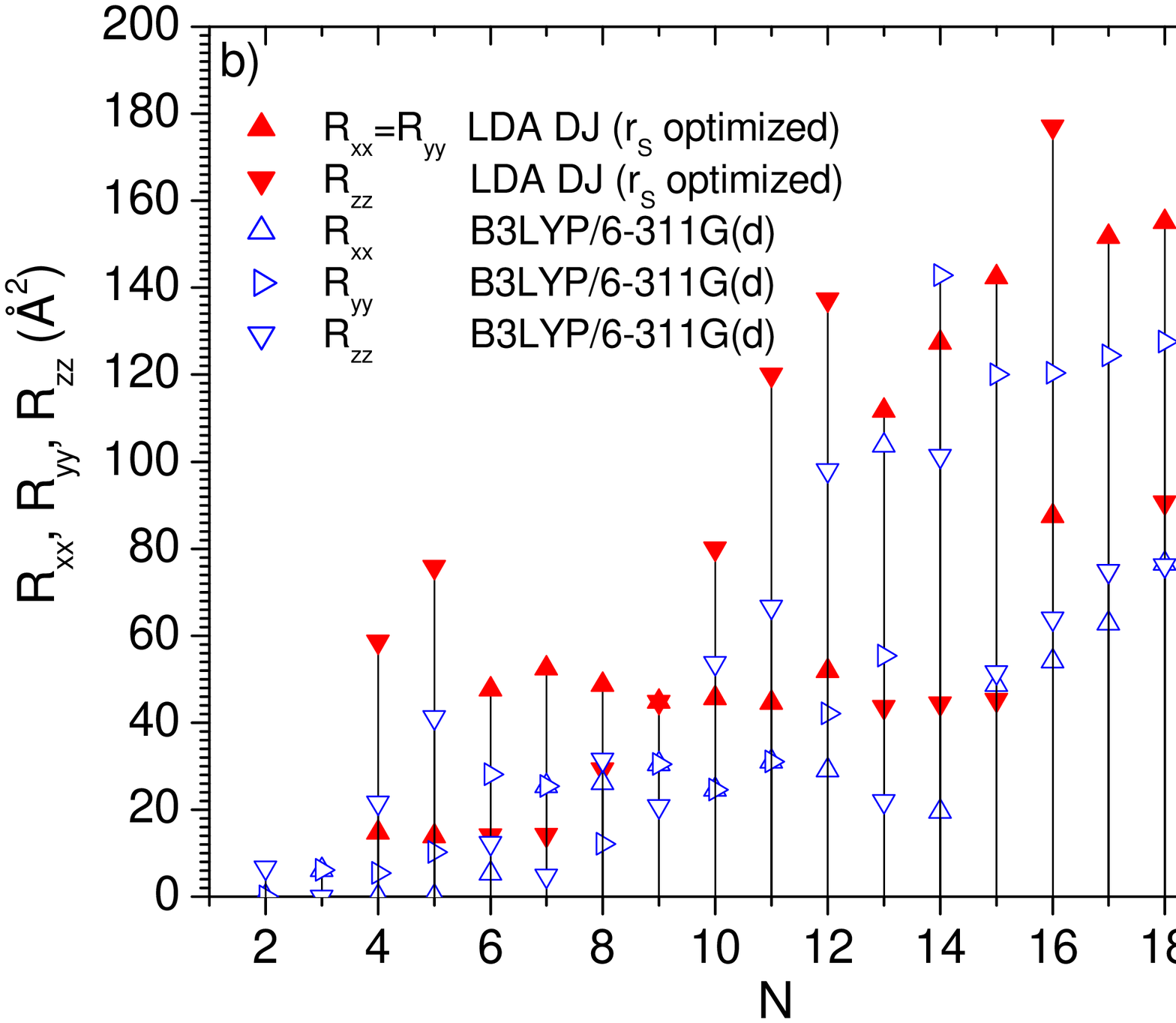}
\end{center}
\caption{
The same as Fig. \ref{Tensor_n}, but
for singly charged sodium clusters.
}
\label{Tensor_i}
\end{figure}

In figure \ref{Tensor_n},  we present the principle values
$R_{xx}$, $R_{yy}$  and $R_{zz}$  calculated for the neutral sodium clusters
with $N< 20$ in the framework of the deformed jellium model according
to (\ref{tensor_R_j}). The diameters $a$ and $b$ have been determined by
minimizing the total cluster energy  
in the LDA   approximation (\ref{E-el-LDA}) as explained in the
previous subsection. This calculation has been performed
with $r_s=4.0$ (Fig. \ref{Tensor_n}a), 
which corresponds to the density of the bulk sodium,
and with optimized value of the Wigner-Seitz
radius (Fig. \ref{Tensor_n}b). The cluster energy minimization on Wiger-Seitz
radius will be discussed in section \ref{WSopt} in more detail. 
The LDA deformed jellium model results 
are shown in figure \ref{Tensor_n} by the filled triangles. 
The filled
triangles pointing up correspond to $R_{xx}$=$R_{yy}$, while
those pointing down to $R_{zz}$. The opened triangles are
the results of the all-electron {\it ab initio} framework derived
in \cite{StructNa} with the use of the B3LYP density functional. 
The opened triangles pointing up and down show 
$R_{xx}$  and $R_{zz}$ respectively, while
the opened triangles pointing  right represent  $R_{yy}$.
In figure \ref{Tensor_i} we present the results of similar 
calculations performed for singly charged ions. 
The notations used in figure \ref{Tensor_i} are the same
as in figure \ref{Tensor_n}.

Comparison of the results presented in 
figures \ref{Tensor_n} and \ref{Tensor_i},  demonstrate
that the optimization of the cluster energy on the Wigner-Seitz
radius does not change much the neutral cluster geometries. For
cluster ions, the alteration of the cluster shape with the
variation of the Wigner-Seitz radius is more noticeable, although
it does not improve the agreement of the jellium model and
{\it ab initio} results. This comparison demonstrates that 
the Wigner-Seitz radius variation in the jellium model does  not
actually improve the quality of the model.

Figures \ref{Tensor_n} and \ref{Tensor_i} demonstrate rather good
agreement of the jellium model and {\it ab initio} results. In most
of the cases the jellium model predicts correctly the type of the dominant
cluster deformation, prolate or oblate one. Of course, {\it ab initio} 
calculations include tri-axial deformations of the cluster, which
turned out to be noticeable for the clusters with the open subshells and play
important role for clusters like $Na_{12}- Na_{14}$, $Na_{17}$,
$Na_{13}^{+}$, $Na_{14}^{+}$, $Na_{20}^{+}$.
The axially symmetric deformed jellium model does not take into account
tri-axial deformations and thus in this case always $R_{xx}=R_{yy}$.

The axially symmetric
jellium model gives  the wrong type of deformation in 
the open shell clusters, like $Na_5$, $Na_{16}- Na_{19}$, 
$Na_6^+ $, $Na_{17}^{+}- Na_{20}^+$. 
However, it is necessary to note that 
for all these clusters there are almost degenerate oblate and prolate isomers
within the axially  symmetric jellium model, 
as it is shown in figures \ref{tot_en_n} and \ref{tot_en_i} and in
tables \ref{tab:neutral} and \ref{tab:ion} (see Appendix). 
Thus, accounting for tri-axial deformations in these clusters
plays the crucial role as it becomes clear from the comparison of the
jellium and {\it ab initio} results.

For the magic clusters $Na_{8}$ and  $Na_{20}$, the principle values 
$R_{xx}=R_{yy}=R_{zz}$ are almost identical in both approaches, which
demonstrates the closeness to the sphericity of the {\it ab initio} 
magic cluster shapes. Note that  for the magic cluster ion, $Na_9^+$, 
some small deformation remains in the {\it ab initio} approach, while
in the jellium approach it turns out to be spherical. This demonstrates
that the ionic framework of the cluster is not that deformable as it
follows from the jellium model. 

In Appendix, we compiled in tables the optimized 
Wigner-Seitz radii $r_{s}$, total energies per
atom $E_{tot}(N,\delta)/N$, deformation parameters $\delta$ 
and the second derivatives of the total energy on 
cluster deformation 
at the $\delta$-point corresponding to minimum of the total energy. The later 
characteristic of the cluster, 
$\partial^2 E_{tot}(N,\delta) / \partial \delta^2$,
is directly connected to the frequency of  
cluster surface vibrations (see, e.g., \cite{GISG00} for details). 
All values which are presented in the 
tables \ref{tab:neutral} and \ref{tab:ion} 
have been calculated in the LDA deformed jellium model 
for neutral and singly charged clusters.

\subsection{Binding energies per atom}

In this paper we calculate the dependence of binding energy per atom
in the deformed jellium model and compare it with {\it ab initio}
results from \cite{StructNa}. The binding energies per atom for the 
neutral and singly charged clusters are defined as follows:
\begin{eqnarray}
E_b/N&=&E_1-E_N/N 
\label{E_b}
\\
E_b^+/N&=& \left((N-1)E_1+E_1^+-E_N^+\right)/N,
\label{E_b^p}
\end{eqnarray}

\noindent
where $E_N$ and $E_N^+$ are the total energies of a neutral  and
singly-charged N-atomic jellium cluster
respectively.

\begin{figure}
\begin{center}
\includegraphics[bb=15 283 828 851, scale=0.5]{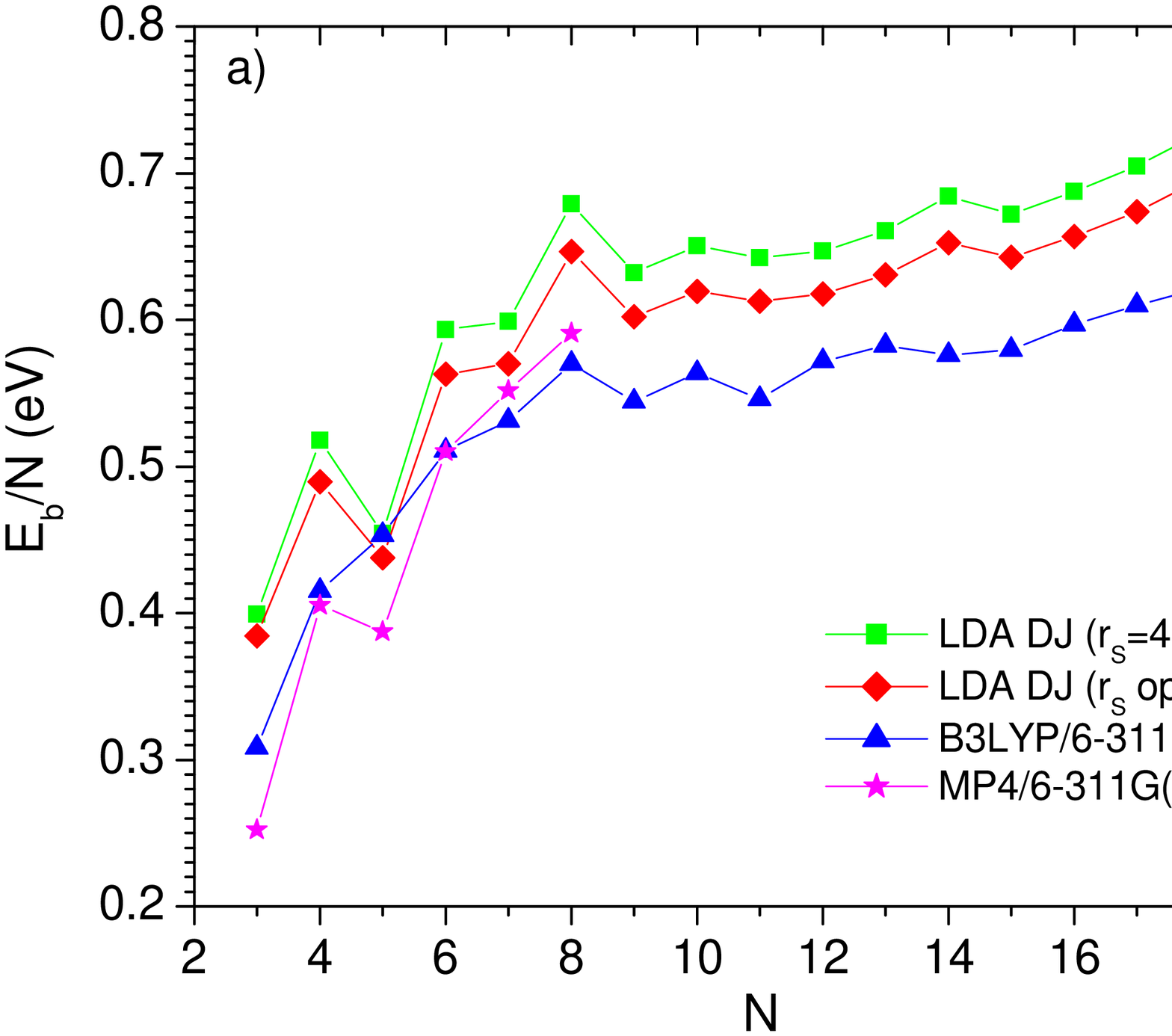}
\includegraphics[bb=15 283 828 851, scale=0.5]{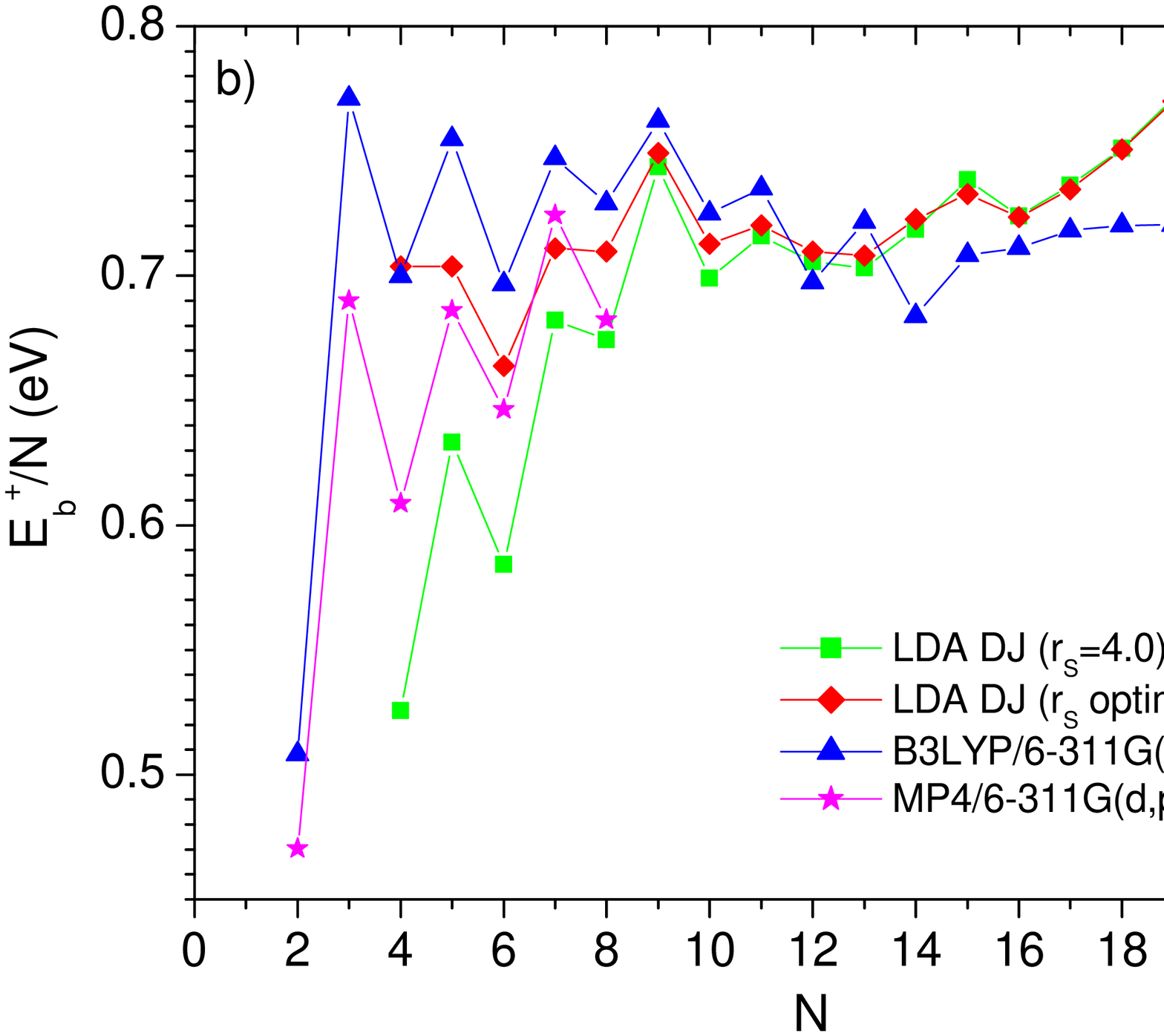}
\end{center}

\caption{
Binding energy per atom for neutral (a) and singly
charged (b) sodium clusters as a function of cluster
size calculated in the LDA deformed jellium model and compared with
{\it ab initio} B3LYP and MP4 results from \cite{StructNa}.
}
\label{Bind_en}
\end{figure}

Figure \ref{Bind_en} shows the dependence of the binding 
energy per atom for neutral (Fig. \ref{Bind_en}a) 
and singly charged (Fig. \ref{Bind_en}b) 
clusters as a function of cluster size 
calculated in the deformed 
jellium model. We compare the calculated dependences with the
{\it ab initio}  results  from \cite{StructNa} obtained
by the  B3LYP and MP4 methods.
In figure \ref{Bind_en}  we  
show the jellium model results obtained with bulk and optimized
values of the Wigner-Seitz radius. It is seen that the cluster optimization
on the Wigner-Seitz radius brings the cluster energies down and 
makes them closer
to the {\it ab initio} results. 

Figure \ref{Bind_en} demonstrates
that the general trend of the curves calculated within the
jellium framework turns out to be very close to the one obtained from
the {\it ab initio} calculation.  The similarity of the
the jellium and {\it ab initio} curves is higher for $N\leq 10$.
In the region $10 \leq N \leq 20$ small discrepancy in the behaviour
of the curves can be attributed to the tri-axial cluster deformations taken
into account in the {\it ab initio} approach and omitted in the
axially symmetric jellium model. 

Note that the jellium model results for both neutral
and singly charged sodium clusters are somewhat closer to
the predictions of the $MP4$ method. This method is based on
the accounting of the many-electron correlations up to the
fourth order of the perturbation theory and is free of
any adjustable parameters. 

Figure \ref{Bind_en} demonstrates
that in spite of the simplicity, the jellium model turns out to be
rather reliable approximation able to reproduce reasonably well
the dependence of binding energy per atom for both neutral
and singly charged sodium clusters.

\subsection{On the role of exchange and many electron correlation interaction.}
\label{correlations}

\begin{figure}
\begin{center}
\includegraphics[bb=15 283 828 851, scale=0.5]{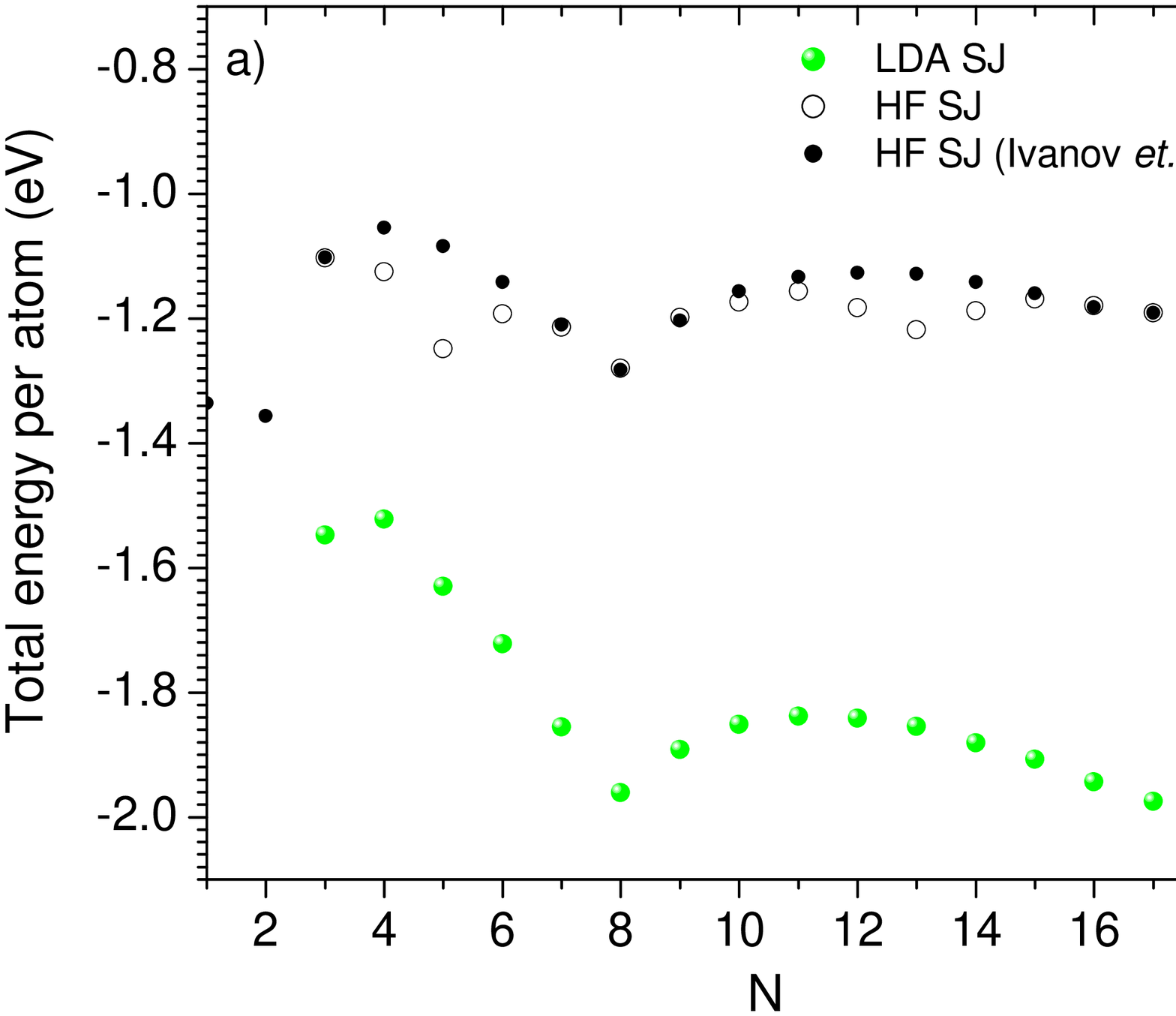}
\includegraphics[bb=15 283 828 851, scale=0.5]{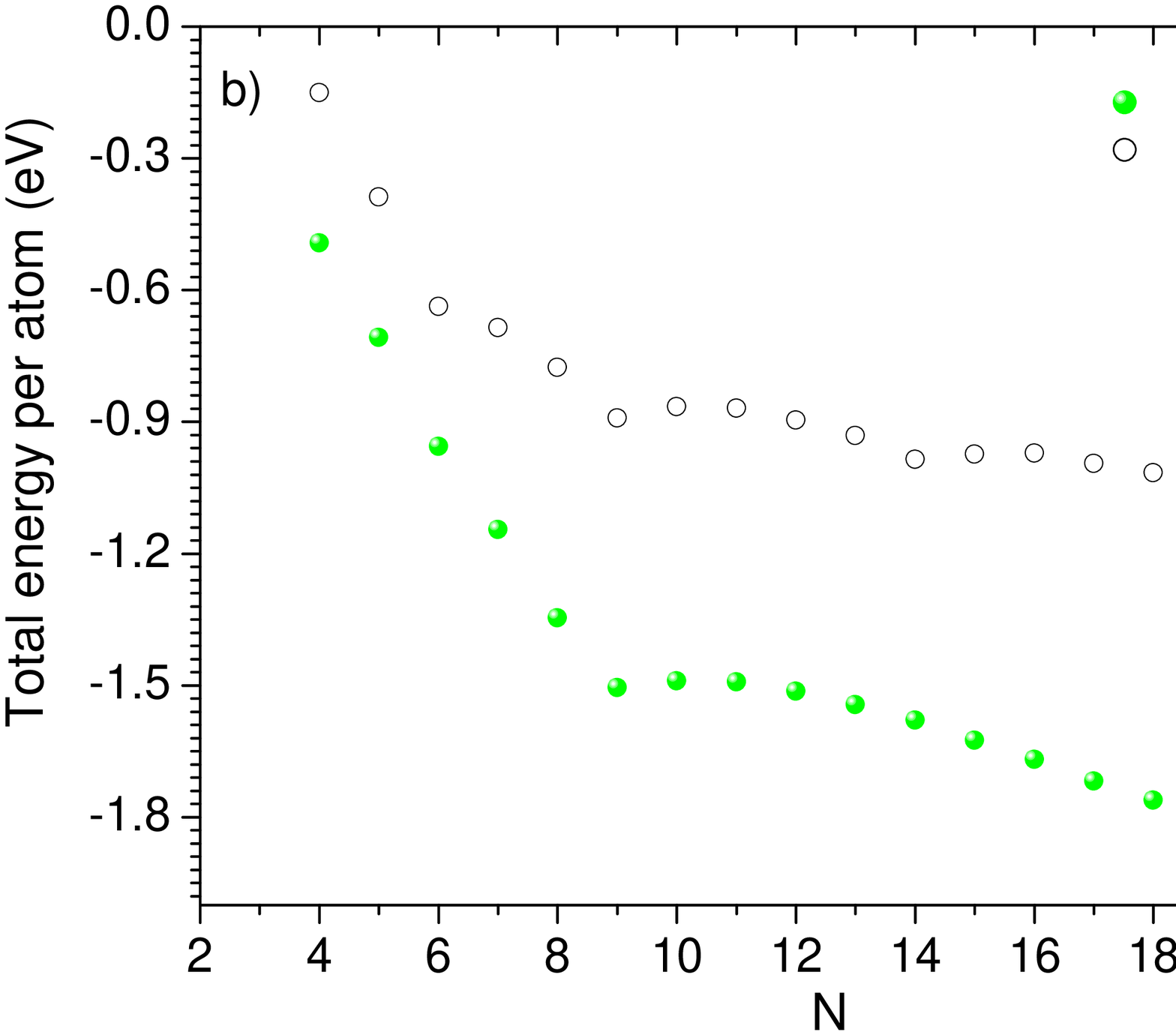}
\end{center}
\caption{
Total energy per atom  for spherical neutral (a) and
and singly charged (b) sodium clusters calculated in 
the Hartree-Fock and LDA spherical jellium model 
(HF SJ and LDA SJ).
}
\label{Spher_cl}
\end{figure}

To illustrate the importance of the exchange and many-electron
correlation interaction in the cluster, we plot on figure \ref{Spher_cl} 
the  total energy per atom calculated  both in the HF and LDA 
spherical jellium approximations
for neutral (Fig. \ref{Spher_cl}a) and singly charged (Fig. \ref{Spher_cl}b) 
spherical sodium clusters as a function of cluster size. 
The spherical cluster shape was assumed
in this calculation for the sake of simplicity.  Figure \ref{Spher_cl}
shows the significant difference between the HF and LDA results for both
neutral and singly charged clusters, which is the result of the accounting
for the many-electron correlation interaction within LDA.

\begin{figure}[p]
\begin{center}
\includegraphics[bb=15 283 828 851, scale=0.5]{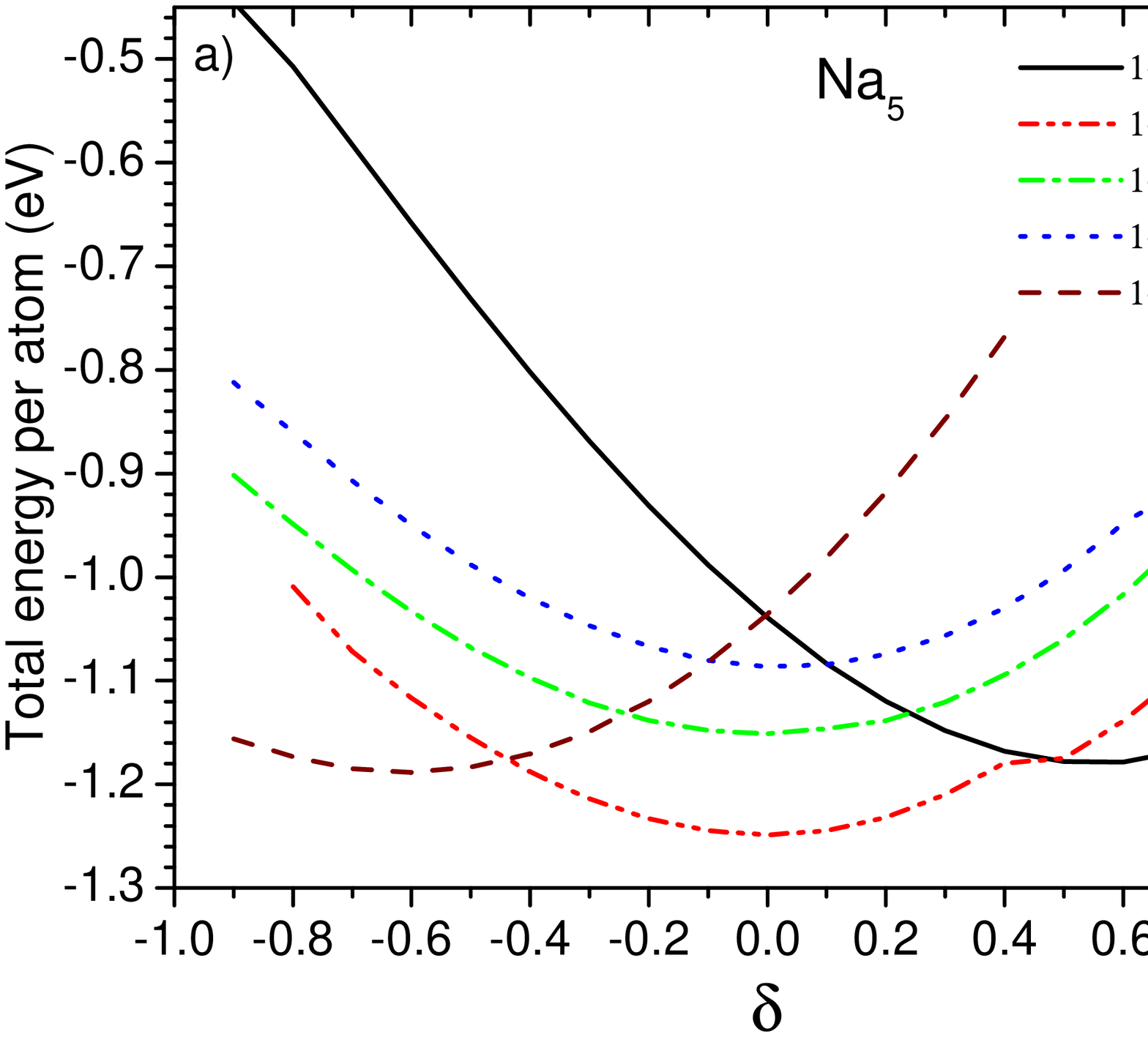}
\includegraphics[bb=15 283 828 851, scale=0.5]{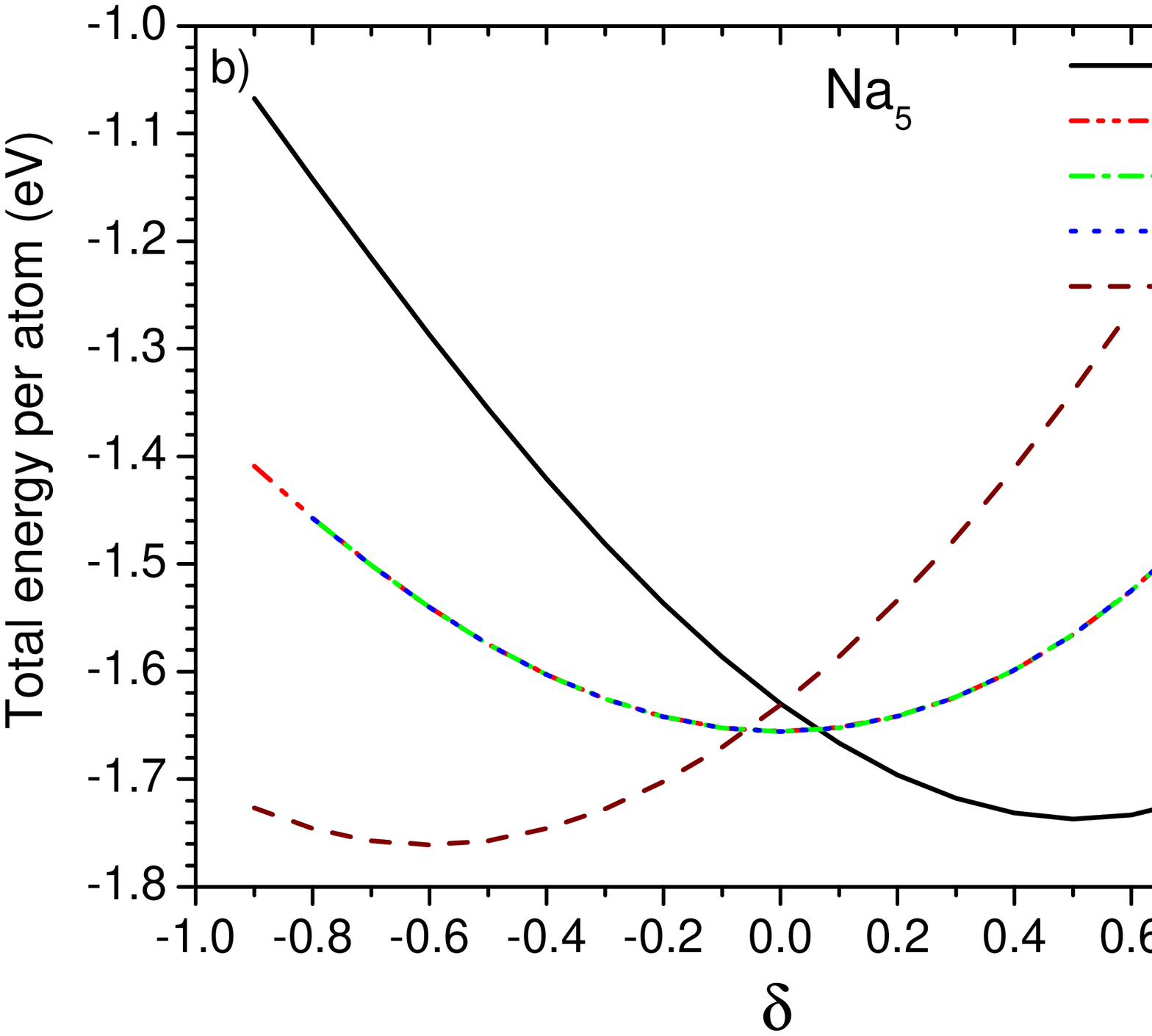}
\end{center}
\caption{Total energy per atom  for $Na_5$ cluster versus deformation parameter 
$\delta$ calculated in the Hartree-Fock (a) and LDA (b) deformed 
jellium model
for different electronic configurations.}
\label{Spin}
\end{figure}

Figure \ref{Spher_cl} demonstrates that the LDA total energy dependence 
possesses the local minima at the shell closings $N= 2, 8, 20$ for
neutral clusters and $N= 3, 9, 21$ for singly charged ones.
The HF total energy curve has the extra minima at the half-shell closings, i.e.
$N= 5, 13, 19$ for neutral clusters and $N= 6, 14, 20$ for singly charged
cluster ions. The appearance of these extra minima is the result of 
the more accurate accounting for the exchange interaction
within the HF approximation. We found that with increasing cluster size
within the given shell the favorable electronic configuration of the cluster
changes resulting in the formation of the extra minima on the 
HF total energy dependence. Qualitatively, this situation can be understood
on the basis of the Hund's rule which states that the lowest energy level 
in the system at a fixed electronic configuration is characterized 
by the maximum value of total spin and the maximum possible (at this
spin) angular momentum. 
As an illustration, figure \ref{Spin} shows the 
$\delta$-dependencies of the total energy per atom for $Na_{5}$ cluster
obtained in the Hartree-Fock (Fig. \ref{Spin}a) and LDA 
(Fig. \ref{Spin}b) deformed jellium model
for different electronic configurations. 
Figure \ref{Spin}a clearly demonstrates that the 
extra minima on HF total energy curve
correspond to the electronic states in which spins of all the electrons
from the open shell are parallel. 
The LDA framework, with the exchange-correlation potential
(\ref{LDA-pot}),
which we have used in
our work does not take into account the spin polarization effects and thus 
we found  no minima on the LDA total energy curve 
at closing of the half filled shells (see Fig. \ref{Spin}b).

\subsection{Ionization potentials}

Another important characteristic of the cluster system
is its ionization potential. The ionization potential is determined by
the energy needed to take an electron out of the cluster. 
It is equal to:

\begin{equation}
V_{i}=E_N^+-E_N 
\label{IP}
\end{equation}

\begin{figure}
\begin{center}
\includegraphics[bb=15 283 828 851, scale=0.5]{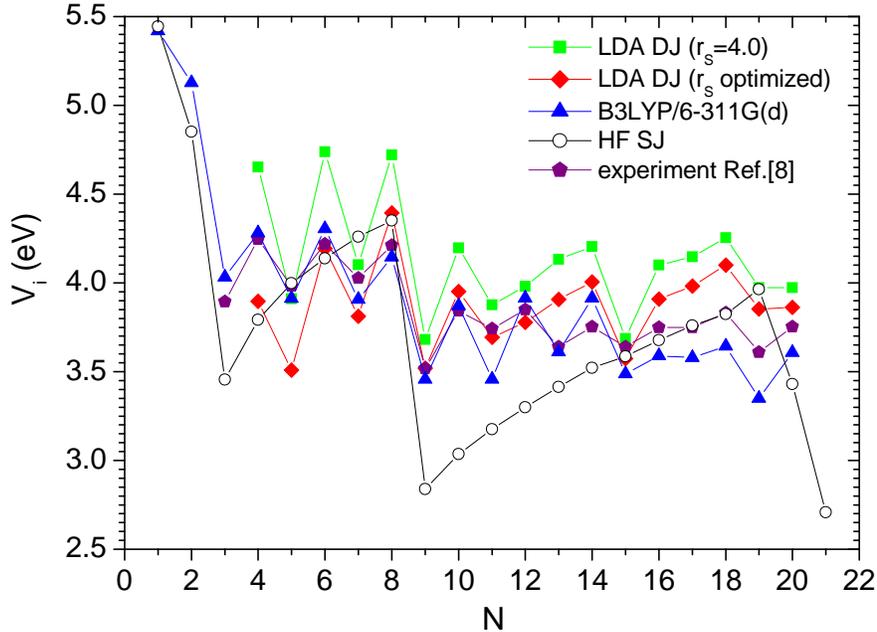}
\end{center}
\caption{
Ionization potentials of neutral sodium clusters calculated
in the deformed jellium model 
and compared with 
{\it ab initio} results from \cite{StructNa} and
with experiment \cite{deHeer93}.
}
\label{ip}
\end{figure}

In figure \ref{ip}, we present the ionization potential of neutral
sodium clusters calculated within the jellium model as a function
of cluster size. We compare the jellium model results with those
obtained in \cite{StructNa}  using {\it ab initio} theoretical framework and
with the available experimental data \cite{deHeer93}. 
This comparison demonstrates that the jellium model reproduces
correctly most of the essential features of the ionization
potential dependence on $N$. Some discrepancy, like in the region
$11\leq N \leq 14$, can be attributed to the neglection of the tri-axial
deformation in the axially symmetric jellium model.

In spite of the fact that {\it ab initio} results are closer to the
experimental points, one can state quite satisfactory agreement of
the jellium model results with the experimental data, which
illustrates correctness of the jellium model assumptions and its
applicability to the description of sodium clusters. 

Figure \ref{ip} also demonstrates the role of cluster deformations
on the formation of the odd-even oscillations in the dependence
of the cluster ionization potential on $N$. Indeed, for spherically
symmetric clusters this dependence turns out to be monotonous 
within the range of the given shell \cite{Ivanov96} contrary
to the experimental observations. Allowing for the cluster deformation
and introducing a single deformation parameter $\delta$, we have achieved
much better agreement of theoretical results with the experimental data as
it is clear from figure \ref{ip}.

\subsection{Wigner-Seitz radius variation}
\label{WSopt}

Calculations of the cluster total energy are usually performed at the certain
value of the Wigner-Seitz radius $r_s$. The bulk value of the Wigner-Seitz
radius for sodium is equal to 4.0. However, 
one can also perform the calculation minimizing the total cluster energy
by variation of the Wigner-Seitz radius.

\begin{figure}
\begin{center}
\includegraphics[bb=15 283 828 851, scale=0.5]{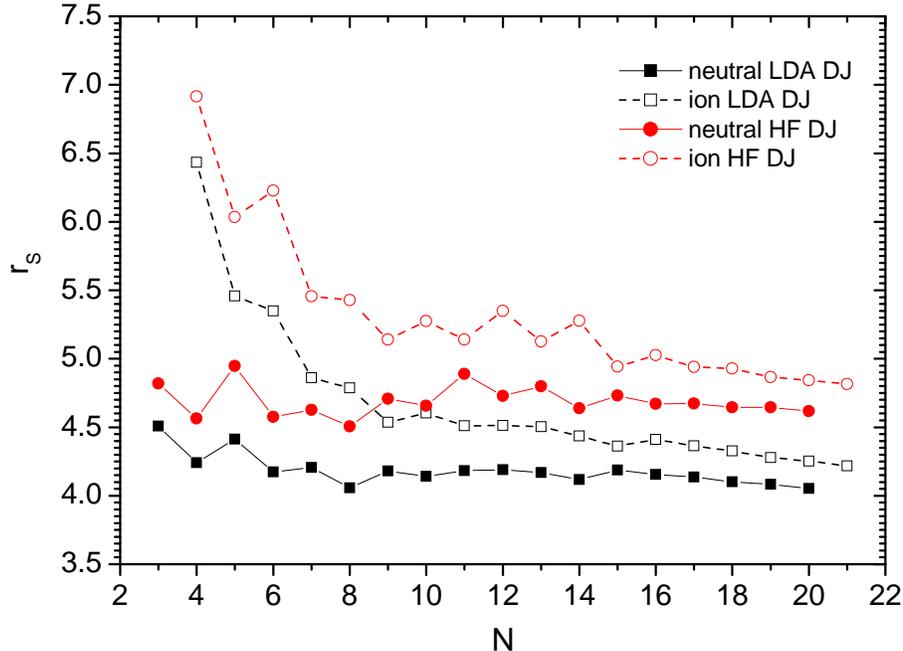}
\end{center}
\caption{Optimized Wigner-Seitz radii for
neutral and singly charged sodium clusters calculated as a function
of cluster size in the HF and LDA deformed jellium models.
}
\label{rs}
\end{figure}

Figure \ref{rs} demonstrates the dependence of the
optimized Wigner-Seitz radii on cluster size calculated
for neutral and singly charged sodium clusters within the HF and LDA
approximations.  This figure shows that the alteration of
the optimized $r_s$ values is much larger for the cluster ions
as compared to the neutral clusters. 
For neutral clusters, the optimized
values are somewhat larger than the bulk value $r_s=4.0$ 
in both LDA and HF approximations. The LDA dependence
goes closer to the bulk limit. With
increasing $N$ this dependence approaches
the bulk limit, being very close to it also for
the magic numbers $N=8$ and $N=20$, which is another
manifestation of the shell effect. 

\begin{figure}
\begin{center}
\includegraphics[bb=15 283 828 851, scale=0.5]{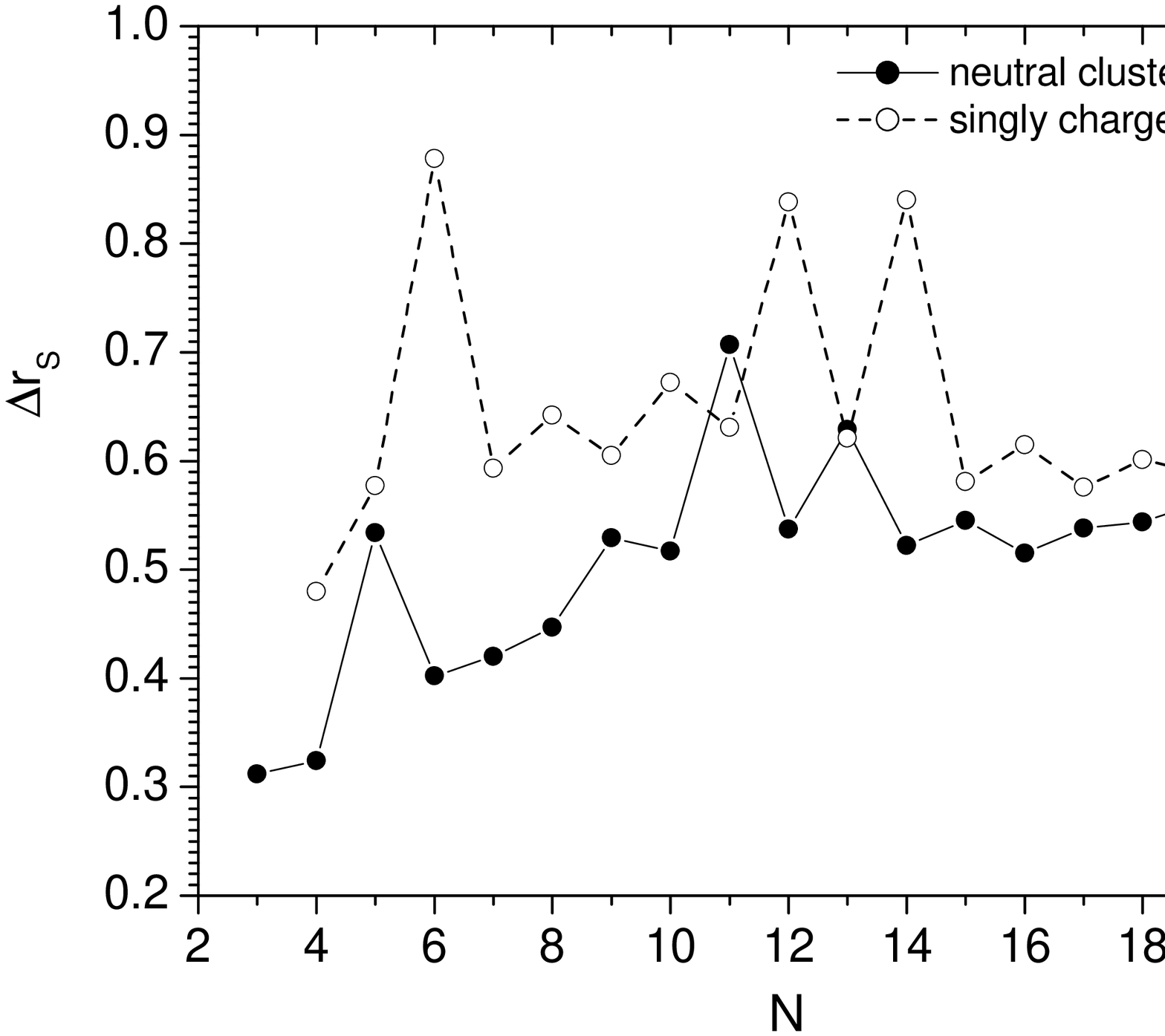}
\end{center}
\caption{Difference between the
optimized values of the Wigner-Seitz radius for
neutral and singly charged sodium clusters calculated as a function
of cluster size in HF and LDA deformed jellium model.
}
\label{rs_diff}
\end{figure}

The difference between the optimized values of the Wigner-Seitz
radii obtained in the HF and LDA approximations can be atributed
to the manifestation of the many-electron correlation interaction
in the system. To illustrate this fact, we plot 
in figure \ref{rs_diff}  the difference
between the optimized Wigner-Seitz radii
calculated in the LDA and HF approximations for neutral
and singly charged sodium clusters.
These dependences have the prominent peculiarities. The origin
of these peculiarities is the same as 
for those in figure \ref{Spher_cl} (see section \ref{correlations} and discussion therein), 
although now we consider deformed cluster
systems.

\section{Conclusion}
\label{conclusion}

In this paper we performed systematic calculation of 
various characteristics of neutral and singly charged sodium 
clusters with $N\leq 20$ 
on the basis of the deformed HF and LDA jellium models.
We compared the results of our calculations with 
the {\it ab initio} results obtained in \cite{StructNa}
and with the available experimental data.
From these comparisons, we have established the level
of applicability of the jellium model to the description of
various cluster characteristics.

Our consideration shows that the deformed jellium model
provides qualitatively correct description of the sodium clusters
and their ions. The quantitatively reliable results with the accuracy
below than 10 per cent one can expect from the jellium model description
providing one allows for the tri-axial cluster deformations.

We have performed our calculations using HF and LDA approximations.
Comparison of the results of the two approaches allowed us to demonstrate
the importance of the many-electron correlations in the formation of
cluster characteristics and properties.

We have performed our calculations for sodium clusters. However,
most of the conclusions should be applicable to other alkali clusters,
potassium for example. The level of applicability of the jellium 
approach to other metals, like alkali-earth, requires a separate
careful consideration.
 
\begin{acknowledgments}
The authors acknowledge support from the Alexander von Humboldt Foundation, 
DAAD and the Russian Academy of Sciences (Grant 44).
\end{acknowledgments}

{

\newpage
\appendix*
\section{Tables}

In Appendix, we present tables of the optimized 
Wigner-Seitz radii, total energies per
atom, deformation parameters and  
the second derivatives of the total energy on 
cluster deformation calculated in the LDA deformed jellium 
model for neutral and singly charged sodium clusters.

\begingroup
\squeezetable  
\begin{table*}[h]
\caption{ Summary of the optimized 
Wigner-Seitz radii $r_{s}$, 
total energies per atom $E_{tot}(N,\delta)/N$, deformation parameters 
$\delta$ and  second derivatives of the total energy on 
cluster deformation, $\partial^2 E_{tot}(N,\delta) / \partial \delta^2$,
calculated at $\delta$ corresponding to the minimum of the total energy.
Calculations have been performed in the LDA 
deformed jellium model for neutral sodium clusters.}
\begin{ruledtabular}
\begin{tabular}{cclccc}  

\multicolumn{1}{c}{N} &   
\multicolumn{1}{c}{$r_{S}$} &   
\multicolumn{1}{c}{Electronic configuration} &  
\multicolumn{1}{c}{$E_{tot}/N$ (eV)} &    
\multicolumn{1}{c}{$\delta$} & 
\multicolumn{1}{c}{$\partial^2 E_{tot}(N,\delta) /\partial\delta^2$} \\    
\hline

 3  & 4.51 & 1$\sigma$2$\sigma^{\uparrow}$   & -1.700 &  0.60 &  0.81 \\
    &      & 1$\sigma$1$\pi^{\uparrow}$      & -1.592 & -0.38 &  0.53 \\

 4  & 4.24 & 1$\sigma$2$\sigma$                              & -1.805 &  0.78 & 1.11\\
    &      & 1$\sigma$2$\sigma^{\uparrow}$1$\pi^{\uparrow}$  & -1.588 &  0.26 & 0.69\\  

 5  & 4.41 & 1$\sigma$1$\pi^{\uparrow\downarrow\uparrow}$    & -1.775 & -0.58 & 0.75\\
    &      & 1$\sigma$2$\sigma$1$\pi^{\uparrow}$             & -1.753 &  0.49 & 0.90\\  

 6  & 4.17 & 1$\sigma$1$\pi$                                 & -1.878 & -0.62 & 0.81\\
    &      & 1$\sigma$2$\sigma$1$\pi^{\uparrow\uparrow}$     & -1.784 &  0.28 & 0.85\\  

 7  & 4.21 & 1$\sigma$2$\sigma^{\uparrow}$1$\pi$                    & -1.885 & -0.26 & 0.77\\
    &      & 1$\sigma$2$\sigma$1$\pi^{\uparrow\downarrow\uparrow}$  & -1.867 &  0.12 & 0.82\\  

 8  & 4.06 & 1$\sigma$2$\sigma$1$\pi$                                                   & -1.962 &  0.00 & 1.02\\
    &      & 1$\sigma$2$\sigma$1$\pi^{\uparrow\downarrow\uparrow}$3$\sigma^{\uparrow}$  & -1.828 &  0.42 & 0.83\\

 9  & 4.18 & 1$\sigma$2$\sigma$1$\pi$3$\sigma^{\uparrow}$           & -1.917 &  0.27 & 0.78\\
    &      & 1$\sigma$2$\sigma$1$\pi$3$\delta^{\uparrow}$           & -1.903 & -0.17 & 0.84\\  

 10 & 4.14 & 1$\sigma$2$\sigma$1$\pi$3$\sigma$                             & -1.935 &  0.49 & 0.97\\
    &      & 1$\sigma$2$\sigma$1$\pi$3$\sigma^{\uparrow}$2$\pi^{\uparrow}$ & -1.890 &  0.31 & 0.88\\  

 11 & 4.18 & 1$\sigma$2$\sigma$1$\pi$3$\sigma$2$\pi^{\uparrow}$                        & -1.928 &  0.48 & 1.03\\
    &      & 1$\sigma$2$\sigma$1$\pi$3$\sigma^{\uparrow}$1$\delta^{\uparrow\uparrow}$  & -1.877 & -0.36 & 0.78\\
      
 12 & 4.19 & 1$\sigma$2$\sigma$1$\pi$3$\sigma$2$\pi^{\uparrow\uparrow}$     & -1.933 &  0.47 & 1.07\\
    &      & 1$\sigma$2$\sigma$1$\pi$1$\delta$	                            & -1.923 & -0.47 & 0.97\\  

 13 & 4.17 & 1$\sigma$2$\sigma$1$\pi$3$\sigma^{\uparrow}$1$\delta$                & -1.946 & -0.51 & 0.93\\
    &      & 1$\sigma$2$\sigma$1$\pi$3$\sigma$2$\pi^{\uparrow\downarrow\uparrow}$ & -1.945 &  0.46 & 1.10\\  

 14 & 4.12 & 1$\sigma$2$\sigma$1$\pi$3$\sigma$1$\delta$                   & -1.968 & -0.56  & 0.93\\
    &      & 1$\sigma$2$\sigma$1$\pi$3$\sigma$2$\pi$	                  & -1.941 &  0.46 & 1.17\\  

 15 & 4.19 & 1$\sigma$2$\sigma$1$\pi$3$\sigma$2$\pi$1$\delta^{\uparrow}$                     & -1.958 &  0.35 & 1.05\\ 
    &      & 1$\sigma$2$\sigma$1$\pi$3$\sigma^{\uparrow}$2$\pi$1$\delta^{\uparrow\uparrow}$  & -1.941 & -0.31 & 0.81\\  

 16 & 4.16 & 1$\sigma$2$\sigma$1$\pi$3$\sigma$2$\pi^{\uparrow\uparrow}$1$\delta$  & -1.972 & -0.35 & 0.80\\
    &      & 1$\sigma$2$\sigma$1$\pi$3$\sigma$2$\pi$1$\delta^{\uparrow\uparrow}$  & -1.964 &  0.26 & 0.98\\  

 17 & 4.14 & 1$\sigma$2$\sigma$1$\pi$3$\sigma$2$\pi^{\uparrow\downarrow\uparrow}$1$\delta$      & -1.989 & -0.27 & 0.76\\
    &      & 1$\sigma$2$\sigma$1$\pi$3$\sigma$2$\pi$1$\delta^{\uparrow\downarrow\uparrow}$      & -1.963 &  0.16 & 0.82\\  

 18 & 4.10 & 1$\sigma$2$\sigma$1$\pi$3$\sigma$2$\pi$1$\delta$           & -2.011 & -0.20  & 0.57\\
    &      & 1$\sigma$2$\sigma$1$\pi$3$\sigma$2$\pi$1$\delta$           & -2.008 &  0.06  & 0.93\\                                                       

 19 & 4.08 & 1$\sigma$2$\sigma$1$\pi$3$\sigma$2$\pi$1$\delta$4$\sigma^{\uparrow}$                     & -2.016 &  0.00  & 0.51\\
    &      & 1$\sigma$2$\sigma$1$\pi$3$\sigma$2$\pi$1$\delta^{\uparrow\downarrow\uparrow}$4$\sigma$   & -1.988 &  0.05  & 1.25\\
 
 20 & 4.05 & 1$\sigma$2$\sigma$1$\pi$3$\sigma$2$\pi$1$\delta$4$\sigma$   & -2.022 &  0.00 & 1.01\\
    &      & 1$\sigma$2$\sigma$1$\pi$3$\sigma$2$\pi$1$\delta$4$\sigma$	 & -1.990 &  0.37 & 1.02\\

\end{tabular}  
\end{ruledtabular}
\label{tab:neutral}
\end{table*}  
\endgroup

\newpage

\begingroup
\squeezetable  
\begin{table*}[h]
\caption{ The same as Tab. \ref{tab:neutral}, but for the 
singly charged sodium cluster ions.}
 
\begin{ruledtabular}
\begin{tabular}{cclccc}  

\multicolumn{1}{c}{N} &   
\multicolumn{1}{c}{$r_{S}$} &   
\multicolumn{1}{c}{Electronic configuration} &  
\multicolumn{1}{c}{$E_{tot}/N$ (eV)} &    
\multicolumn{1}{c}{$\delta$} & 
\multicolumn{1}{c}{$\partial^2 E_{tot}(N,\delta) /\partial\delta^2$} \\   
\hline

 4  &	6.43 &	1$\sigma$2$\sigma^{\uparrow}$                     & -0.831 &  0.74 & 0.63\\
    &	     &	1$\sigma$2$\pi^{\uparrow}$                        & -0.544 & -0.50 & 0.36\\
 
 5  &	5.46 &	1$\sigma$2$\sigma$	                          & -1.073 &  0.86 & 0.89\\
    &        &  1$\sigma$2$\sigma^{\uparrow}$1$\pi^{\uparrow}$    & -0.771 &  0.28 & 0.56\\
 
 6  &	5.35 &	1$\sigma$1$\pi^{\uparrow\downarrow\uparrow}$      & -1.179 & -0.59 & 0.62\\
    &        &  1$\sigma$2$\sigma$1$\pi^{\uparrow}$		  & -1.159 &  0.54 & 0.77\\

 7  &	4.86 &	1$\sigma$1$\pi$	                                  & -1.341 & -0.67  & 0.71\\
    &        &  1$\sigma$2$\sigma$1$\pi^{\uparrow\uparrow}$	  & -1.203 &  0.29  & 0.74\\

 8  &	4.79 &  1$\sigma$2$\sigma^{\uparrow}$1$\pi$	              & -1.412 & -0.27  & 0.69\\
    &        &  1$\sigma$2$\sigma$1$\pi^{\uparrow\downarrow\uparrow}$ & -1.358 &  0.12  & 0.73\\
                
 9  &	4.54 &  1$\sigma$2$\sigma$1$\pi$	                                          & -1.526 &  0.00  & 0.76\\
    &	     &  1$\sigma$2$\sigma$1$\pi^{\uparrow\downarrow\uparrow}\sigma^{\uparrow}$   & -1.387 & -0.44  & 0.80\\
 
 10 &	4.60 &	1$\sigma$2$\sigma$1$\pi$3$\sigma^{\uparrow}$         & -1.539 &  0.29  & 0.73\\
    &        &  1$\sigma$2$\sigma$1$\pi$3$\delta^{\uparrow}$         & -1.501 & -0.17  & 0.78\\

 11 &	4.51 &	1$\sigma$2$\sigma$1$\pi$3$\sigma$                             & -1.592 &  0.50 & 0.92\\
    &        &  1$\sigma$2$\sigma$1$\pi$3$\sigma^{\uparrow}$2$\pi^{\uparrow}$ & -1.529 &  0.32 & 0.81\\

 12 &	4.51 &	1$\sigma$2$\sigma$1$\pi$3$\sigma$2$\pi^{\uparrow}$                       & -1.618 &  0.49 & 0.96\\
    &        &  1$\sigma$2$\sigma$1$\pi$3$\sigma^{\uparrow}$1$\delta^{\uparrow\uparrow}$ & -1.551 & -0.37 & 0.72\\

 13 &	4.50 &  1$\sigma$2$\sigma$1$\pi$3$\sigma$2$\pi^{\uparrow\uparrow}$  & -1.645 & -0.48 & 1.01\\
    &        &  1$\sigma$2$\sigma$1$\pi$1$\delta$			    & -1.648 &  0.48 & 0.90\\

 14 &	4.44 &  1$\sigma$2$\sigma$1$\pi$3$\sigma^{\uparrow}$1$\delta$                & -1.682 & -0.53 & 0.87\\
    &        &  1$\sigma$2$\sigma$1$\pi$3$\sigma$2$\pi^{\uparrow\downarrow\uparrow}$ & -1.681 &  0.47 & 1.04\\

 15 &	4.36 &	1$\sigma$2$\sigma$1$\pi$3$\sigma$1$\delta$                          & -1.719 & -0.57 & 0.87\\
    &        &  1$\sigma$2$\sigma$1$\pi$3$\sigma$2$\pi$ 		            & -1.701 &  0.46 & 1.09\\

 16 &	4.41 &	1$\sigma$2$\sigma$1$\pi$3$\sigma$2$\pi$1$\delta^{\uparrow}$                    & -1.728 &  0.36 & 1.00\\
    &        &  1$\sigma$2$\sigma$1$\pi$3$\sigma^{\uparrow}$2$\pi$1$\delta^{\uparrow\uparrow}$ & -1.700 & -0.31 & 0.76\\

 17 &	4.37 &  1$\sigma$2$\sigma$1$\pi$3$\sigma$2$\pi^{\uparrow\uparrow}$1$\delta$          & -1.755 & -0.36 & 0.78\\
    &        &  1$\sigma$2$\sigma$1$\pi$3$\sigma$2$\pi$1$\delta^{\uparrow\uparrow}$	     & -1.737 &  0.26 & 0.93\\

 18 &	4.33 &	1$\sigma$2$\sigma$1$\pi$3$\sigma$2$\pi^{\uparrow\downarrow\uparrow}$1$\delta$ & -1.783 & -0.27 & 0.66\\
    &        &  1$\sigma$2$\sigma$1$\pi$3$\sigma$2$\pi$1$\delta^{\uparrow\downarrow\uparrow}$ & -1.768 &  0.16 & 0.74\\

 19 &	4.28 &  1$\sigma$2$\sigma$1$\pi$3$\sigma$2$\pi$1$\delta$                                                   & -1.814 & -0.20  & 0.44\\
    &        &  1$\sigma$2$\sigma$1$\pi$3$\sigma$2$\pi$1$\delta^{\uparrow\downarrow\uparrow}$4$\sigma^{\uparrow}$  & -1.782 &  0.08  & 0.89\\

 20 &	4.25 &	1$\sigma$2$\sigma$1$\pi$3$\sigma$2$\pi$1$\delta$4$\sigma^{\uparrow}$                    & -1.829 &  0.00  & 0.50\\
    &	     &	1$\sigma$2$\sigma$1$\pi$3$\sigma$2$\pi$1$\delta^{\uparrow\downarrow\uparrow}$4$\sigma$  & -1.795 &  0.05  & 1.20\\
 
 21 &	4.22 &	1$\sigma$2$\sigma$1$\pi$3$\sigma$2$\pi$1$\delta$4$\sigma$   & -1.816 &  0.00 & 0.97\\
    &        &  1$\sigma$2$\sigma$1$\pi$3$\sigma$2$\pi$1$\delta$4$\sigma$   & -1.812 &  0.37 & 0.96\\

\end{tabular}  
\end{ruledtabular}
\label{tab:ion}
\end{table*} 
\endgroup

}

\end{document}